\documentclass[aps, prd, preprint, groupedaddress, nofootinbib]{revtex4-1}
\usepackage{amsmath, bbold, dsfont, rotating, hyperref, tikz, pgfplots, footmisc}
\usepackage{subfig}
\usepackage{pgf}
\begin{document}
\title{Initial Systematic Investigations of the Landscape of Low Layer NAHE Extensions}
\author{Timothy Renner}
\email{Renner.Timothy@gmail.com}
\author{Jared Greenwald}
\email{Jared\_Greenwald@baylor.edu}
\author{Douglas Moore}
\email{Douglas\_Moore1@baylor.edu}
\author{Gerald Cleaver}
\email{Gerald\_Cleaver@baylor.edu}
\affiliation{Baylor University}
\date{\today}

\begin{abstract}
The discovery that the number of physically consistent string vacua is on the order of $10^{500}$ has prompted several statistical studies of string phenomenology.
Contained here is one such study that focuses on the Weakly Coupled Free Fermionic Heterotic String (WCFFHS) formalism.
Presented are systematic extensions of the well-known NAHE (Nanopoulos, Antoniadis, Hagelin, Ellis) set of basis vectors, which have been shown to produce phenomenologically realistic models.
Statistics related to the number of $U(1)$'s, the specific gauge groups and their factors, non-Abelian singlets, and spacetime supersymmetries (ST SUSYs) are discussed for the full range of models produced.
These findings are compared with prior results of other large-scale investigations.
Statistical coupling between the gauge groups and the number of  ST SUSYs is also discussed. In particular, for order-3 extensions there is a correlation between the appearance of exceptional groups and enhanced ST SUSY.
Also discussed are some three-generation GUT models found among the data sets.
These models are unique because they come from basis vectors which still have a geometric interpretation -- there are no ``rank-cuts" in these models.
\end{abstract}
\preprint{BU-HEPP-11-06}
\maketitle

\section{Introduction}
Physically consistent string/M-theory derived models number at least approximately $10^{500}$ \cite{Bousso:2000, Ashok:2003}.
The large number of vacua has prompted both computational and analytical examinations of the the ``landscape" of possible string vacua, e.g.\ 
\cite{Dijkstra:2004,Donagi:2004, Valandro:2008, Balasubramanian:2008, Lebedev:2008, Gmeiner:2008, Dienes:2008, Gabella:2008, Donagi:2008}.
The Weakly Coupled Free Fermionic Heterotic String (WCFFHS)\cite{Antoniadis:1986, Antoniadis:1987, Kawai:1986_2,Kawai:1988} approach to string model construction has produced some of the most phenomenologically realistic string models to date
\cite{Cleaver:1999, Lopez:1992, Faraggi:1989, Faraggi:1992, Antoniadis:1990, Leontaris:1999, Faraggi:1991, Faraggi:1992_2, Faraggi:1992_3, Faraggi:1991_2, Faraggi:1991_3, Faraggi:1995, Faraggi:1996, Cleaver:1997, Cleaver:1997_2, Cleaver:1997_3, Cleaver:1998, Cleaver:1998_2, Cleaver:1998_3, Cleaver:1998_4, Cleaver:1998_5, Cleaver:1999_2, Cleaver:1999_3, Cleaver:1999_4, Cleaver:2000, Cleaver:2000_2, Cleaver:2001, Cleaver:2001_2, Cleaver:2002, Cleaver:2002_2, Cleaver:2002_3, Perkins:2003, Perkins:2005, Cleaver:2008, Greenwald:2009, Faraggi:2010a,Gato-Rivera:2010a,Gato-Rivera:2010b,Gato-Rivera:2010c,Gato-Rivera:2011a,Faraggi:2011a,Cleaver:2011}.
Outlined in this section are the basic steps needed to construct models within this formalism.
Also detailed are the methodology of systematic searches and the difficulties they pose, and an explanation of the NAHE set of basis vectors used to construct realistic string models.
Finally, results of the preliminary order-2 and order-3 extension to the NAHE set are presented, with an emphasis on GUT group models.

\subsection{Weakly Coupled Free Fermionic Heterotic String Model Building}
The heterotic theories consist of closed strings which have independent sets of left and right moving modes. 
The left moving modes are ten dimensional superstrings, while the right moving modes are 26 dimensional bosonic strings. 
WCFFHS models are constructed by specifying the phases that fermion modes gain when parallel transported around non-contractible loops on a genus-1 world sheet of a string. 
The number of possible values these phases may take is referred to as the order, while the number of ``basis vectors" of phases used to specify the model is referred to as the layer. 
The linear combinations of the basis vectors produce sectors from which the physical states of the model are built. 
The force and matter content is then determined from the symmetries of these states.

Modular invariance highly constrains the possible components of the basis vectors.
A consistent  set of basis vectors $\{\vec{\alpha}^B\}$ is such that:
\begin{eqnarray}
N_{ij}\vec{\alpha}^B_i\cdot \vec{\alpha}^B_j&=&0\pmod{4},\\
N_{i}\vec{\alpha}^B_i\cdot \vec{\alpha}^B_i&=&0\pmod{8} ({\text{for even\  } N_i}),
\end{eqnarray}
where $N_{ij}$ is the least common multiple of the orders of the basis vectors $\vec{\alpha}^B_{i}$ and 
$\vec{\alpha}^B_{j}$, and the dot product is Lorentz,
with a relative minus sign between the left-moving and the right-moving contributions.
Additionally,
\begin{center}
\textit{The number of simultaneous periodic modes for any three basis vectors must be even.}
\end{center}
Taking all possible linear combinations of the basis vectors builds the full ``geometry" of the model, designated by sectors $\{\vec{\alpha}\}$.
States from a sector $\vec{\alpha}$ are designated by their charge vectors $\vec{Q}_{\vec{\alpha}} $, defined by 
\begin{equation}
\vec{Q}_{\vec{\alpha}} = \frac{1}{2}\vec{\alpha} + \vec{F},
\end{equation}
where $\vec{F}$ is all combinations of 0,$\pm$1. The massless (with regard to the Planck scale) states are such that
\begin{eqnarray}
\vec{Q}_L^2=2,\\
\vec{Q}_R^2=4,
\end{eqnarray}
in a real fermion basis. 
In a complex basis, the values are halved.
A physical state in a model a state must survive all GSO projections (GSOPs) from basis vectors $\vec{\alpha}^B_j$.
These require that
\begin{equation}
\vec{\alpha}^B_j\cdot\vec{Q}_{\vec{\alpha}} = \sum_i a_ik_{ji} + s_j\pmod{2},
\end{equation}
where $a_i$ are the coefficients on the basis vectors used to construct $\vec{\alpha}$, $s_j$ is 0 (1) for bosonic (fermionic) basis vectors 
$a^B_j$ , and $k_{ji}$ is an element of the GSO coefficient matrix.

The GSO  matrix must also follow its own modular invariance constraints, given by 
\begin{eqnarray}
N_j k_{ij} &=&0\pmod{2},\\
k_{ij} + k_{ji} &=& \frac{1}{2} \vec{\alpha}^B_i \cdot \vec{\alpha}^B_j\pmod{2},\quad {\rm and}\\
k_{ii} + k_{i1} &=& \frac{1}{4} \vec{\alpha}^B_i \cdot \vec{\alpha}^B_i - s_i\pmod{2}.
\end{eqnarray}

\subsection{Challenges to Systematic Searches}
Systematic searches are designed to do a ``complete" scan of the possible consistent sets of basis vectors and GSO coefficients that define models within a chosen parameter space.
In the discussions herein, models are classified into the order (the number of available twists for right-moving modes) and the layer (the number of basic vectors).
Searches with random sampling have already been performed \cite{Dienes:2006, Dienes:2007_2, Dienes:2007,Assel:2010}.
However, random sampling methods have several difficulties which are non-trivial to address \cite{Dienes:2007}.
Systematic methods avoid these inherent difficulties because models are not ``sampled" - they form a complete set.
Described in this section are the motivations and challenges to systematic searches of WCFFHS models including model building speed, model data storage, and model comparison.

\subsection{The Scale of Systematic WCFFHS Searches}
The scope of the systematic WCFFHS model searches can be limited by several factors.
The first, and most obvious, factor is that of model building speed.
Because the number of models to be built can get quite large, minimization of construction speed is essential.
Assuming the model building process takes around 1 second, a data set with 1,000,000 models would take about 28 hours.
As the number of ``layers" of basis vectors in a model increases, the number of total models to build grows geometrically; each of the 1,000,000 models now has on the order of 1,000,000 new basis vectors (likely more than that) with modular invariance.
Thus, the total number of models in the data set of the next layer is an estimated $10^{12}$. 
At one model every second, the layer-2 data set would be completed in approximately 3000 years.

Several steps can be taken to reduce the time required for model construction, both from an analytical and implementation standpoint.
From the implementation side, understanding optimization techniques when writing the model building software is essential.
Minimizing file and screen i/o, as well as the amount of copying done in memory are crucial steps needed to reduce the computing time.

Another computational concern is one of data storage.
While a ``master atlas" of basis vectors and the models to which they map is desirable, it is not feasible.
The estimated average amount of space per model is 0.725 kB.
For the 1,000,000 model data set, this amounts to 725 MB.
The next layer would use approximately $725\times10^3$ TB, making a pure ``atlas" of WCFFHS models unreasonable to pursue.
There are two solutions to this problem.
One is to gather statistics as the models are generated without keeping the models themselves in memory.
While this approach is less taxing on system resources, it does have disadvantages.
In particular, one must know the statistics to be gathered on the models prior to runtime.
Any additional statistics would require a second run.
Moreover, the total model set would have to be statistically analyzed, as opposed to the distinct models only, as the models are not being kept in memory.
Double-counts that result from the construction method itself cannot be avoided with this approach.
The other approach is to count the distinct models only, writing those to a file that serves as a small repository.
While a full discussion of uniqueness in WCFFHS models is discussed in the next section, it suffices to say there are disadvantages to this approach as well.
In particular, the definition of uniqueness in these models can be somewhat nebulous at times. 
Additionally, each unique model that is found must be compared with the other unique models that have been constructed, making each model require more computing time to complete.

Analytic techniques can also be used to reduce the number of redundant models produced.
As these techniques are related to uniqueness in WCFFHS phenomenology, discussion of such analysis will be deferred until after uniqueness has been addressed.

\subsection{Uniqueness in WCFFHS Models}\label{sec: Uniqueness_in_WCFFHS_Models}
The previously described computational limitations force the analysis to be only on models that are considered distinct.
However, the definition of distinctness amongst WCFFHS models is not always clear.
In particular the ``amount" of phenomenology done to distinguish the models from one another must be balanced with the amount of computing time required to perform such analyses, as well as how easily the analyses can be automated.

Consider two different basis vector sets producing the same gauge and non-Abelian matter content.
Even though the actual charge vectors and gauge group eigenvalues for the two models may be different, the overarching group structures are identical.
This could be due to a string-scale type symmetry within the construction method itself, in which case there would be some transformation that could be applied to one of the basis vector sets to reproduce the other.
The $U(1)$ charges of those two models, once diagonalized, would be the same.
Such a symmetry in the construction method could in principle be revealed through an analytic study.
However, it could also be the case that these models have non-equivalent $U(1)$ structures, and that the GSOPs eliminated the states that were different between the models.
Such models will not have identical superpotentials or D- and F-flat directions, all of which have significant impact on the phenomenological viability of a model.
For the systematic searches in this study the $U(1)$ charges of the matter representations will not be considered (although they will in future studies). 
Thus, two models will be considered identical herein  if they have the same gauge and non-Abelian matter content and the same number of non-Abelian singlets.

This approach has a caveat, however.
As the models become increasingly complex, comparing two models becomes increasingly difficult to automate.
Consider the following two ``toy" models whose particle content is presented in Tables \ref{tab: Toy_Models}.
\begin{table}
\caption{Two models illustrating the inherent difficulty in comparing WCFFHS models.}
\begin{center}
\begin{tabular}{||c|c|c|c|c|c||}
\hline \hline
QTY&$SU(4)$&$SU(4)$&$SU(4)$&$SO(10)$&$E_8$\\
\hline \hline
1&4&4&1&1&1\\
\hline
1&4&1&4&1&1\\
\hline
1&1&4&4&1&1\\
\hline
2&1&6&1&10&1\\
\hline
2&1&1&6&10&1\\
\hline \hline
\end{tabular}
\vspace{3mm}\\
\begin{tabular}{||c|c|c|c|c|c||}
\hline \hline
QTY&$SU(4)$&$SU(4)$&$SU(4)$&$SO(10)$&$E_8$\\
\hline \hline
1&4&4&1&1&1\\
\hline
1&4&1&4&1&1\\
\hline
1&1&4&4&1&1\\
\hline
2&6&1&1&10&1\\
\hline
2&1&1&6&10&1\\
\hline\hline
\end{tabular}
\end{center}
\label{tab: Toy_Models}
\end{table}
It is clear that these two models are equivalent if two of the $SU(4)$ groups are switched.
However, a simple boolean comparison of the gauge groups and matter states by a computer program will result in these models being counted as distinct.
The root of the problem lies in the fact that the three $SU(4)$ gauge groups are not identical --- there are different matter representations that transform under them.
The most obvious solution would be to perform brute force permutations on the identical gauge groups and resorting the matter representations.
Such an approach would work, but takes a significant amount of computing time.
Comparisons are the most called operation in a systematic search, however, and thus need to be as fast as possible in order for the search to be efficient.

The solution implemented in this study is to propose and use a conjecture that defines uniqueness in a slightly stronger way.
The conjecture is that identical models will have matter that fits into the same ``classes" of representations.
These matter representation classes are formed by ignoring the singlets (after verifying equal numbers of singlets in models 
under cpmparison) and considering only at the dimension of the non-Abelian representations and the gauge groups under which they transform.
The classes of matter representations for the two example models are presented in Table \ref{tab: Toy_Model_Classes}.
\begin{table}
\caption{Matter representation classes of the ``toy" models.}
\begin{center}
\begin{tabular}{|ccc|}
\hline \hline
QTY&Model 1 Classes & Model 2 Classes\\
\hline \hline
\multicolumn{3}{|c|}{$SU(4),SU(4)$}\\
\hline\hline
1&(4,4)&(4,4)\\
\hline
1&(4,4)&(4,4)\\
\hline
1&(4,4)&(4,4)\\
\hline\hline
\multicolumn{3}{|c|}{$SU(4),SO(10)$}\\
\hline\hline
2&(6,10)&(6,10)\\
\hline
2&(6,10)&(6,10)\\
\hline \hline
\end{tabular}
\end{center}
\label{tab: Toy_Model_Classes}
\end{table}
Now it is clear the two models are likely equivalent. This would imply that modular invariance  prevents models that have only one representation switched from those of a known modular invariant model.

\begin{table}
\caption{Another ``toy" model, declared non-existant by the conjecture about matter representation classes.}
\begin{center}
\begin{tabular}{||c|c|c|c|c|c||}
\hline \hline
QTY&$SU(4)$&$SU(4)$&$SU(4)$&$SO(10)$&$E_8$\\
\hline \hline
2&4&4&1&1&1\\
\hline
1&4&1&4&1&1\\
\hline
2&1&6&1&10&1\\
\hline
2&1&1&6&10&1\\
\hline\hline
\end{tabular}
\end{center}
\label{tab: Impossible_Toy_Model}
\end{table}

By this conjecture, the modular invariance of Table \ref{tab: Toy_Models}  models would imply the Table 
\ref{tab: Impossible_Toy_Model}  model is not modular invariant.
More theoretical work will need to be done to prove or disprove this conjecture.

The conjecture does not fully remedy the problem of a preferred gauge group ordering.
Generally, matter representations are limited by the charges they carry due to the masslessness constraints of the fermion states.
Representations of larger dimension tend to carry smaller dimensional charges under the other groups in the model, if at all.
Certain gauge groups produce representations of similar dimension, however, which still place an ordering on the gauge groups within a class of matter representation.
In particular, the 4- and 5-dimensional representations of $SO(5)$, the 4- and 6- dimensional representations of $SU(4)$, and the 2- and 3-dimensional representations of $SU(2)^{(2)}$ can still cause double counting amongst models.
To completely remove the dependence on ordering in WCFFHS models, the problem must be reduced to a counting problem.
In addition to proving or disproving the conjecture, the following possibility could also be explored:
Two models with identical gauge groups and ST SUSYs are identical if they have the same total number of distinct matter representation classes, the same number of distinct matter representation classes, and the same number of total fermions.

This would remove the ordering dependence from the model comparison.
Until it is proven, however, there is a risk of undercounting the models.
Rather than risking this undercounting with the systematic studies presented herein, models in smaller data sets were examined, and duplicates were removed by hand.
This will provide a systematic uncertainty estimate for statistics from larger data sets.

\section{The NAHE Set}\label{sec: The_NAHE_Set}
The NAHE \cite{Antoniadis:1989,Faraggi:1992} set is a set of five order-2 basis vectors which have served as a common basis set for phenomenologically realistic WCFFHS models. 
These basis vectors are given in Table \ref{tab: NAHE_Set_BVs}.
\begin{table}
\caption{The basis vectors and GSO coefficients of the NAHE set arranged into sets of matching boundary conditions. $N_R$ is the order of the right mover. The elements $\psi$, $\overline{\psi}^{~i}$, $\overline{\eta}^{~i}$, and $\overline{\phi}^{~i}$ are expressed in a complex basis, while $x^i$, $y^i$, $w^i$, $\overline{y}^{~i}$, and $\overline{w}^{~i}$ are expressed in a real basis.}
\begin{center}
\begin{tabular}{||c|c|c|c|c|c|c|c|c|c|c||}
\hline \hline
Sec&$N_R$&$\psi$&$x^{12}$&$x^{34}$&$x^{56}$&$\overline{\psi}^{~1,...,5}$&$\overline{\eta}^{~1}$&$\overline{\eta}^{~2}$&$\overline{\eta}^{~3}$&$\overline{\phi}^{~1,...,8}$\\
\hline \hline
$\vec{\mathds{1}}$&2&1&1&1&1&1,...,1&1&1&1&1,...,1\\
\hline
$\vec{S}$&2&1&1&1&1&0,...,0&0&0&0&0,...,0\\
\hline
$\vec{b}_1$&2&1&1&0&0&1,...,1&1&0&0&0,...,0\\
\hline
$\vec{b}_2$&2&1&0&1&0&1,...,1&0&1&0&0,...,0\\
\hline
$\vec{b}_3$&2&1&0&0&1&1,...,1&0&0&1&0,...,0\\
\hline \hline
\end{tabular}
\vspace{5 mm}\\
\begin{tabular}{||c|c|c|c|c||}
\hline \hline
Sec&$N_R$&$y^{~1,2}w^{~5,6}||\overline{y}^{~1,2}\overline{w}^{~5,6}$&$y^{~3,...,6}||\overline{y}^{~3,...,6}$&$w^{~1,...,4}||\overline{w}^{~1,...,4}$\\
\hline \hline
$\vec{\mathds{1}}$&2&1,...,1$~||~$1,...,1&1,...,1$~||~$1,...,1&1,...,1$~||~$1,...,1\\
\hline
$\vec{S}$&2&0,...,0$~||~$0,...,0&0,...,0$~||~$0,...,0&0,...,0$~||~$0,...,0\\
\hline
$\vec{b}_1$&2&0,...,0$~||~$0,...,0&1,...,1$~||~$1,...,1&0,...,0$~||~$0,...,0\\
\hline
$\vec{b}_2$&2&1,...,1$~||~$1,...,1&0,...,0$~||~$0,...,0&0,...,0$~||~$0,...,0\\
\hline
$\vec{b}_3$&2&0,...,0$~||~$0,...,0&0,...,0$~||~$0.,,,.0&1,...,1$~||~$1,...,1\\
\hline
\hline
\end{tabular}
\begin{center}
$k_{ij}$ = 
$\left(\begin{tabular}{c|ccccc}
&$\vec{\mathds{1}}$&$\vec{S}$&$\vec{b}_1$&$\vec{b}_2$&$\vec{b}_3$\\
\hline
$\vec{\mathds{1}}$&1&0&1&1&1\\
$\vec{S}$&0&0&0&0&0\\
$\vec{b}_1$&1&1&1&1&1\\
$\vec{b}_2$&1&1&1&1&1\\
$\vec{b}_3$&1&1&1&1&1\\
\end{tabular}\right)$
\end{center}
\label{tab: NAHE_Set_BVs}
\end{center}
\end{table}
The massless particle spectrum is given in Table \ref{tab: NAHE_Set_Particles}.
\begin{table}
\caption{The matter content of the model produced by the NAHE set.}
\begin{center}
\begin{tabular}{||c|c|c|c|c|c||}
\hline \hline
\textbf{QTY}&$SU(4)$&$SU(4)$&$SU(4)$&$SO(10)$&$E_8$\\
\hline \hline
2&$\overline{4}$&$1$&$1$&$16$&$1$\\
\hline
2&$1$&$\overline{4}$&$1$&$16$&$1$\\
\hline
2&$1$&$1$&$\overline{4}$&$16$&$1$\\
\hline
2&$1$&$1$&$4$&$16$&$1$\\
\hline
1&$1$&$1$&$6$&$10$&$1$\\
\hline
2&$1$&$4$&$1$&$16$&$1$\\
\hline
1&$1$&$6$&$1$&$10$&$1$\\
\hline
1&$1$&$6$&$6$&$1$&$1$\\
\hline
2&$4$&$1$&$1$&$16$&$1$\\
\hline
1&$6$&$1$&$1$&$10$&$1$\\
\hline
1&$6$&$1$&$6$&$1$&$1$\\
\hline
1&$6$&$6$&$1$&$1$&$1$\\
\hline \hline
\end{tabular}
\label{tab: NAHE_Set_Particles}
\end{center}
\end{table}
The NAHE set particle content, in addition to the particles listed in Table \ref{tab: NAHE_Set_Particles}, contains an N=1 ST SUSY. 
The observable sector of the NAHE set is an $SO(10)\times SU(4)^3$ GUT group with three sixteen dimensional $SO(10)$ matter representations serving as the matter generations. 
Each generation is charged under a different $SU(4)$ gauge group, and there are two copies of each representation. 
There are two distinct representations with $SO(10)$ charge 16 and $SU(4)$ charge 4, since the $SU(4)$ charge has a barred and unbarred representation. 
This brings the total number of copies for each generation up to four. 
In addition, the dimension of the $SU(4)$ charge itself is counted as being a set of copies of each generation, so the total number of copies of each generation is sixteen. 
There are no matter representations charged under the $E_8$ group, so it is the designated hidden sector for this model. 
Extensions of the NAHE set do not necessarily keep the designated observable sectors in the model. 
A model could, rather than break down the $SO(10)$ gauge group, break the $E_8$ into an $E_6$, producing an $E_6$ observable sector. 
In the past, when models were constructed individually, this was not common. 
Most individually constructed NAHE based models broke the $SO(10)$ gauge group into a Pati-Salam ($SU(4)\otimes SU(2)\otimes SU(2)$) group, $SU(5)\otimes U(1)$, or the MSSM gauge group ($SU(3)\otimes SU(2)\otimes U(1)$). 
In a systematic search that is no longer guaranteed, since the basis vectors are not organized by the researcher, but rather by the computer program.
The notation for the next sections will change slightly; from here on the term layer will refer to the number of basis vectors after the initial set of five NAHE vectors rather than the total number of basis vectors. 
\clearpage
\section{Statistics for Order-2 Layer-1}\label{sec: NAHE_O2L1_Statistics}
The first set of statistics to be reported here are for extensions to the NAHE set with a single basis vector of order 2. (Thus, extensions of this class can at best reduce the number of copies of each generation from 16 to 8.)
The GSO coefficients were fixed for the NAHE set to those presented in Table \ref{tab: NAHE_Set_BVs}.
The GSO coefficients for the extended basis vector were systematically generated such that all possible combinations consistent with modular invariance were built.
This study was repeated for the NAHE set without the ST SUSY generating basis vector $\vec{S}$, to determine the effect of ST SUSY on the models produced.
\subsection{With $\vec{S}$}
There were 439 unique models produced out of 1,945,088 total models.
Approximately 9.5\% of the models in the data set without rank-cuts were duplicates, and 13\% of the models with rank-cuts were duplicates.
All duplicates were removed prior to the statistical analysis.
The frequency of the individual groups appearing in the unique models is presented in Table \ref{tab: NAHE_O2L1_Gauge_Groups}.
\begin{table}
\caption{The frequency of the individual gauge groups amongst the unique models for the NAHE + O2L1 data set. 
	Gauge groups at Ka$\check{c}$-Moody level higher than 1 are denoted with a superscript indicating the Ka$\check{c}$-Moody level.}
\begin{center}
\begin{tabular}{||c|c|c||}
\hline \hline
Gauge Group & Number of Unique Models & \% of Unique Models \\
\hline \hline
$SU(2)$&365&83.14\%\\
\hline
$SU(2)^{(2)}$&73&16.63\%\\
\hline
$SU(4)$&338&76.99\%\\
\hline
$SU(6)$&2&0.4556\%\\
\hline
$SU(8)$&2&0.4556\%\\
\hline
$SO(5)$&155&35.31\%\\
\hline
$SO(8)$&141&32.12\%\\
\hline
$SO(10)$&160&36.45\%\\
\hline
$SO(12)$&2&0.4556\%\\
\hline
$SO(14)$&3&0.6834\%\\
\hline
$SO(16)$&147&33.49\%\\
\hline
$SO(18)$&1&0.2278\%\\
\hline
$SO(20)$&2&0.4556\%\\
\hline
$SO(22)$&1&0.2278\%\\
\hline
$SO(24)$&1&0.2278\%\\
\hline
$SO(26)$&1&0.2278\%\\
\hline
$E_6$&1&0.2278\%\\
\hline
$E_7$&142&32.35\%\\
\hline
$E_8$&144&32.8\%\\
\hline
$U(1)$&332&75.63\%\\
\hline \hline
\end{tabular}
\label{tab: NAHE_O2L1_Gauge_Groups}
\end{center}
\end{table}
The first item of note is how many models retain at least one of the original gauge groups from the NAHE set. 
Approximately $77\%$ of the models kept at least one $SU(4)$ gauge group, while $\approx~36\%$ of the models kept their $SO(10)$ and $\approx~33\%$ kept their $E_8$. 
The most common gauge group in this set is $SU(2)$, which is expected, since it
is the lowest rank non-Abelian gauge group.
About $17\%$ have an $SU(2)^{(2)}$ gauge group. 
In these models, this happens when a left moving mode is paired with a right moving mode.
As mentioned earlier, left-right paired elements reduce the rank of the gauge lattice of the model and, hence, are referred to as rank cuts.
The non-simply laced gauge group $SO(5)$ also appears due to rank cuts.

Also of interest is the number of models with a $U(1)$ gauge group, which for this data set is quite high.
$U(1)$'s can be problematic when dealing with deeper phenomenology in a model.
The more $U(1)$'s present in a model the more likely the model is to have anomalous charge.
This anomalous charge must be dealt with by finding D- and F-flat directions of the potential.
However, the more $U(1)$ charges present, the more flat directions a model is likely to have, enabling more flexibility when giving mass to observable sector charged exotics.
Discussion of anomalous $U(1)$ charges and flat directions of these models will not be reserved for future papers.

Though most of the models have smaller individual gauge group components, some models have gauge group enhancements.
There are some models with $SO(18)$, $SO(20)$, $SO(22)$, $SO(24)$, and $SO(26)$ gauge groups.
Those groups have rank 9, 10, 11, 12, and 13, respectively, making them higher rank than any one of the NAHE set gauge groups.
This occurs when an added basis vector bridges the gap between the mutually orthogonal sets of states of the original five basis vectors, unifying the root spaces of the individual groups into one larger group.
For order-2 models, however, it is clear this is not common.

\begin{figure}
\begin{center}
\begin{tikzpicture}
\begin{axis} [ybar, ylabel = Number of Distinct Models, xlabel = Number of Gauge Group Factors]
\addplot[draw=black, fill=black]coordinates{
(4,2) (5,36) (6,49) (7,74) (8,68) (9,87) (10,48) (11,36) (12,23) (13,11) (14,5) };
\end{axis}
\end{tikzpicture}
\caption{The number of gauge group factors for each model in the NAHE + O2L1 data set.}
\label{fig: NAHE_O2L1_Gauge_Group_Factors}
\end{center}
\end{figure}
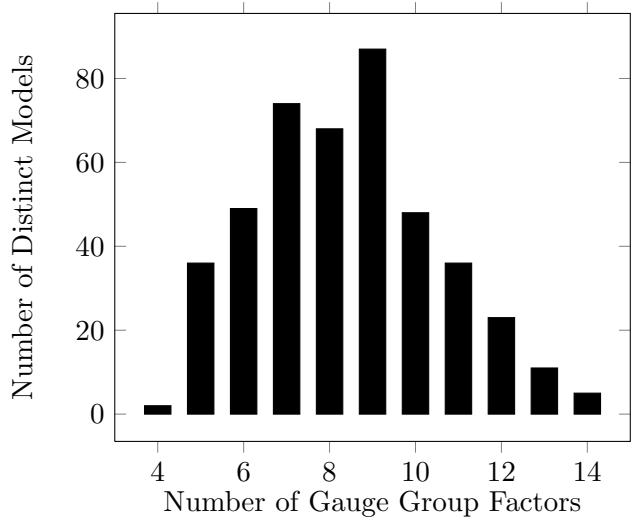

Another way of measuring the enhancements that occur is looking at the number of gauge group factors in each model. 
Those are plotted in Figure \ref{fig: NAHE_O2L1_Gauge_Group_Factors}.
There is a definite peak at nine gauge group factors, much higher than the five initial gauge groups present in the NAHE set.
Note that there are not many models with fewer than five gauge group factors.
This implies that though there are several enhanced groups of higher rank than the initial NAHE set, the other gauge groups in the model remain broken.

Relevant GUT groups and the number of unique models containing those groups is presented in Table \ref{tab: NAHE_O2L1_GUT_Groups}.
\begin{table}
\caption{The number of unique models containing GUT groups for the NAHE + O2L1 data set.}
\begin{center}
\begin{tabular}{||c|c|c||}
\hline \hline
GUT Group & Number of Unique Models & \% of Unique Models \\
\hline \hline
$E_6$ &1&0.2278\%\\
\hline
$SO(10)$&160&36.45\%\\
\hline
$SU(5)\otimes U(1)$&0&0\%\\
\hline
$SU(4)\otimes SU(2)\otimes SU(2)$&243&55.35\%\\
\hline
$SU(3)\otimes SU(2)\otimes SU(2)$&0&0\%\\
\hline
$SU(3)\otimes SU(2)\otimes U(1)$&0&0\%\\
\hline \hline
\end{tabular}
\label{tab: NAHE_O2L1_GUT_Groups}
\end{center}
\end{table}
The relatively low number of $SU(n+1)$ groups (excluding $SU(4)$) explains the lack of GUT group models in this data set.

\begin{figure}
\begin{center}
\begin{tikzpicture}
\begin{axis} [nodes near coords, ybar, ylabel = Number of Distinct Models, xlabel = Number of ST SUSYs]
\addplot[draw=black, fill=black]coordinates{
(0,223) (1,215) (2,1) (4, 0) };
\end{axis}
\end{tikzpicture}
\caption{The number of ST SUSYs for the NAHE + O2L1 data set.}
\label{fig: NAHE_O2L1_ST_SUSYs}
\end{center}
\end{figure}
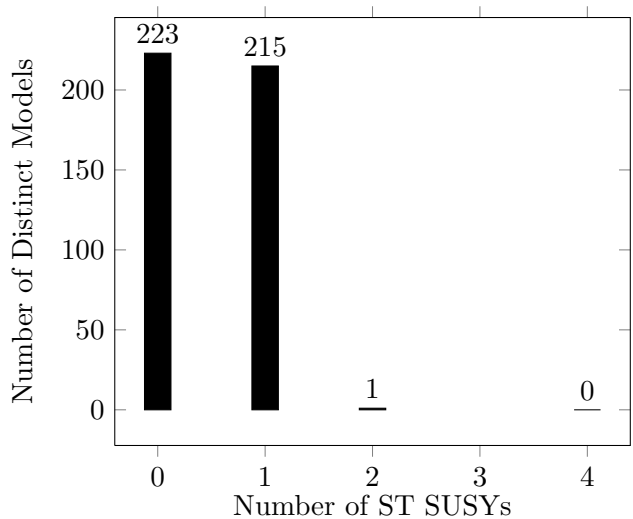

The number of ST SUSYs are plotted against the number of unique models in Figure \ref{fig: NAHE_O2L1_ST_SUSYs}.
Many of the basis vector extensions did not alter the ST SUSY, while about half reduced it. 
More interestingly, there is one model with enhanced ST SUSY.
The basis vector for that model is presented in Table \ref{tab: NAHE_O2_L1_N2_ST_SUSY}.
\begin{table}
\caption{A basis vector and $k_{ij}$ matrix row which produces an enhanced ST SUSY when added the NAHE set.}
\begin{center}
\begin{tabular}{||c|c|c|c|c|c|c|c|c|c|c||}
\hline \hline
Sec&O&$\psi$&$x^{12}$&$x^{34}$&$x^{56}$&$\overline{\psi}^{~1,...,5}$&$\overline{\eta}^{~1}$&$\overline{\eta}^{~2}$&$\overline{\eta}^{~3}$&$\overline{\phi}^{~1,...,8}$\\
\hline \hline
$\vec{v}$&2&1&1&1&1&0,...,0&1&1&0&0,...,0\\
\hline \hline
\end{tabular}
\vspace{5 mm}\\
\begin{tabular}{||c|c|c|c|c||}
\hline \hline
Sec&O&$y^{~1,2}w^{~5,6}||\overline{y}^{~1,2}\overline{w}^{~5,6}$&$y^{~3,...,6}||\overline{y}^{~3,...,6}$&$w^{~1,...,4}||\overline{w}^{~1,...,4}$\\
\hline \hline
$\vec{v}$&2&0,0,1,1$||$1,1,1,1&0,0,1,1$||$1,1,1,1&0,...,0$||$0,...,0\\
\hline \hline
\end{tabular}
\vspace{3 mm}\\
$k_{\vec{v},j}$ = (0, 0, 0, 1, 1)
\label{tab: NAHE_O2_L1_N2_ST_SUSY}
\end{center}
\end{table}
The particle content of this model is presented in Table \ref{tab: NAHE_O2_L1_N2_ST_SUSY_Particles}.
\begin{table}
\caption{The particle content of the $N=2$ ST SUSY NAHE based model.}
\begin{center}
\begin{tabular}{||c|c|c|c|c|c||}
\hline \hline
\textbf{QTY}&$SU(4)$&$SU(4)$&$SU(4)$&$SO(10)$&$E_8$\\
\hline \hline
1&$1$&$\overline{4}$&$1$&$\overline{16}$&$1$\\
\hline
1&$1$&$\overline{4}$&$1$&$16$&$1$\\
\hline
1&$1$&$1$&$\overline{4}$&$\overline{16}$&$1$\\
\hline
1&$1$&$1$&$\overline{4}$&$16$&$1$\\
\hline
1&$1$&$1$&$4$&$\overline{16}$&$1$\\
\hline
1&$1$&$1$&$4$&$16$&$1$\\
\hline
1&$1$&$4$&$1$&$\overline{16}$&$1$\\
\hline
1&$1$&$4$&$1$&$16$&$1$\\
\hline
2&$1$&$6$&$6$&$1$&$1$\\
\hline
2&$6$&$1$&$1$&$10$&$1$\\
\hline \hline
\end{tabular}
\label{tab: NAHE_O2_L1_N2_ST_SUSY_Particles}
\end{center}
\end{table}
The gauge groups of this model are identical to those of the NAHE set, but with fewer matter representations, particularly with regard to the $SU(4)$ charges.
The enhanced ST SUSY comes from a new gravitino generating sector $\vec{b}_1+\vec{b}_2+\vec{v}$, which contributes a single gravitino state to the model.
The other gravitino comes from $\vec{S}$. 
This example highlights the importance of systematic searches; the enhanced ST SUSYs come from very specific basis vectors that combine with the NAHE set to provide unexpected phenomenology.
Though there are not a statistically significant number of models with this property, these models can highlight subtleties in the WCFFHS formulation that may go unnoticed in a random search.

\begin{figure}
\begin{center}
\begin{tikzpicture}
\begin{axis} [ybar, ylabel = Number of Distinct Models, xlabel = Number of $U(1)$ Factors]
\addplot[draw=black, fill=black]coordinates{
(0,107) (1,164) (2,106) (3,48) (4,14) };
\end{axis}
\end{tikzpicture}
\caption{The number of $U(1)$ factors for the NAHE + O2L1 data set.}
\label{fig:NAHE_O2L1_U1_Factors}
\end{center}
\end{figure}
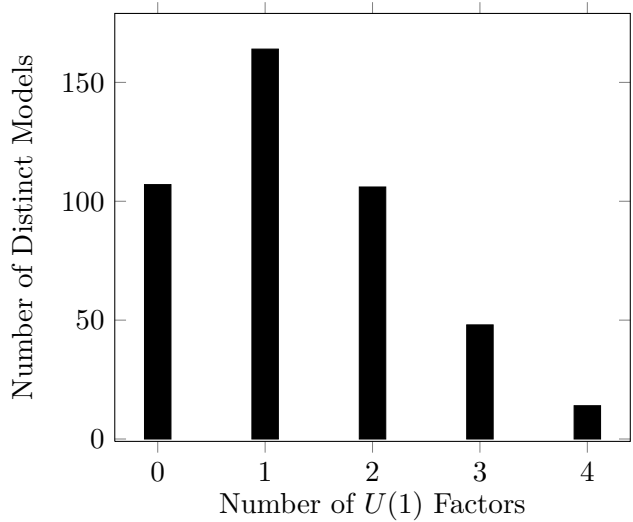

The number of $U(1)$ gauge groups are plotted against the number of unique models in Figure \ref{fig:NAHE_O2L1_U1_Factors}.
The greater number of $U(1)$ factors present in the model, the greater that model's capacity for carrying anomalous charge.
There are relatively few $U(1)$'s in the models of this class.
This is likely the result of the basis vectors having only periodic phases --- nonzero, non-periodic phases break $SO(2n)$ groups into $SU(n-1)\otimes U(1)$ groups in most cases.
Thus, there are not many $U(1)$ groups expected or found in this data set.

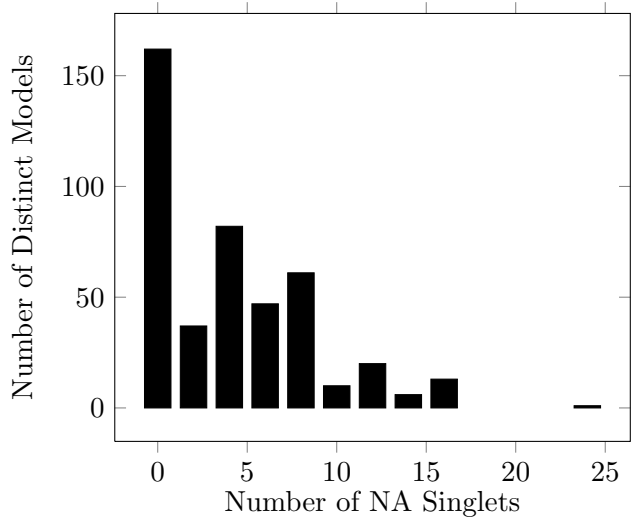
\begin{figure}
\begin{center}
\begin{tikzpicture}
\begin{axis} [ybar, ylabel = Number of Distinct Models,  xlabel = Number of NA Singlets]
\addplot[draw=black, fill=black]coordinates{
(0,162)(2,37)(4,82)(6,47)(8,61)(10,10)(12,20)(14,6)(16,13)(24,1)};
\end{axis}
\end{tikzpicture}
\caption{The number of non-Abelian singlets in the NAHE + O2L1 data set.}
\label{fig:NAHE_O2L1_NA_Singlets}
\end{center}
\end{figure}

Another phenomenological property which might have statistical significance is the number of non-Abelian singlets, as non-Abelian singlets often carry observable sector hypercharge. 
However, no such particle with this property has yet been observed.
Additionally, singlet particles contribute to the mass-energy density of the model. 
Too many non-Abelian singlets could result in a mass-energy density higher than observed values, also producing bad phenomenology.
The number of non-Abelian singlets for this data set is plotted in Figure \ref{fig:NAHE_O2L1_NA_Singlets}.

\subsection{Without $\vec{S}$}
Also of interest is the effect of the $\vec{S}$ vector in the set of NAHE basis vectors. 
Not only does the $\vec{S}$ vector generate ST SUSY, it also adds a degree of freedom to the $k_{ij}$ matrix.
It stands to reason that removing $\vec{S}$ will also have an effect on the massless gauge and matter content in addition to the supersymmetry.
There are 282 unique models in the set of 1,940,352 consistent models, in contrast to the 439 models in the data set with $\vec{S}$. 
About 8.4\% of the models without rank cuts had duplicates, while 9.4\% of the models with rank cuts were duplicates of other models produced.
All duplicates were removed from the statistics to follow.
Moreover the 282 models in this set do not all belong to the $N=0$ models in Figure \ref{fig: NAHE_O2L1_ST_SUSYs}.
In fact, the two data sets have only 8 models in common, implying that the presence or non-presence of $\vec{S}$ has an effect on more than the ST SUSY.
		
The gauge content of the order-2 layer-1 NAHE extensions without the $\vec{S}$ vector are presented in Table \ref{tab: NAHE_NSV_O2L1_Gauge_Groups}.
\begin{table}
\caption{The gauge content of the NAHE + O2L1 data set without $\vec{S}$.}
\begin{center}
\begin{tabular}{||c|c|c||}
\hline \hline
Gauge Group & Number of Unique Models & \% of Unique Models \\
\hline \hline
$SU(2)$&233&82.62\%\\
\hline
$SU(2)^{(2)}$&67&23.76\%\\
\hline
$SU(4)$&212&75.18\%\\
\hline
$SU(6)$&2&0.7092\%\\
\hline
$SU(8)$&2&0.7092\%\\
\hline
$SO(5)$&107&37.94\%\\
\hline
$SO(8)$&89&31.56\%\\
\hline
$SO(10)$&100&35.46\%\\
\hline
$SO(12)$&2&0.7092\%\\
\hline
$SO(14)$&3&1.064\%\\
\hline
$SO(16)$&83&29.43\%\\
\hline
$SO(18)$&1&0.3546\%\\
\hline
$SO(20)$&2&0.7092\%\\
\hline
$SO(22)$&1&0.3546\%\\
\hline
$SO(24)$&1&0.3546\%\\
\hline
$SO(26)$&1&0.3546\%\\
\hline
$E_6$&1&0.3546\%\\
\hline
$E_7$&95&33.69\%\\
\hline
$E_8$&98&34.75\%\\
\hline
$U(1)$&209&74.11\%\\
\hline \hline
\end{tabular}
\label{tab: NAHE_NSV_O2L1_Gauge_Groups}
\end{center}
\end{table}
A side-by-side comparison of the gauge group content percentages is shown in Table \ref{tab: NAHE_O2L1_Gauge_Comparison}.
\begin{table}
\caption{A side-by-side comparison of the gauge content for NAHE + O2L1 with and without $\vec{S}$.}
\begin{center}
\begin{tabular}{||c|c|c||}
\hline \hline
Gauge Group & With $\vec{S}$ & Without $\vec{S}$\\
\hline \hline
$SU(2)$&83.14\%&82.62\%\\
\hline
$SU(2)^{(2)}$&16.63\%&23.76\%\\
\hline
$SU(4)$&76.99\%&75.18\%\\
\hline
$SU(6)$&0.4556\%&0.7092\%\\
\hline
$SU(8)$&0.4556\%&0.7092\%\\
\hline
$SO(5)$&35.31\%&37.94\%\\
\hline
$SO(8)$&32.12\%&31.56\%\\
\hline
$SO(10)$&36.45\%&35.46\%\\
\hline
$SO(12)$&0.4556\%&0.7092\%\\
\hline
$SO(14)$&0.6834\%&1.064\%\\
\hline
$SO(16)$&33.49\%&29.43\%\\
\hline
$SO(18)$&0.2278\%&0.3546\%\\
\hline
$SO(20)$&0.4556\%&0.7092\%\\
\hline
$SO(22)$&0.2278\%&0.3546\%\\
\hline
$SO(24)$&0.2278\%&0.3546\%\\
\hline
$SO(26)$&0.2278\%&0.3546\%\\
\hline
$E_6$&0.2278\%&0.3546\%\\
\hline
$E_7$&32.35\%&33.69\%\\
\hline
$E_8$&32.8\%&34.75\%\\
\hline
$U(1)$&75.63\%&74.11\%\\
\hline \hline
\end{tabular}
\label{tab: NAHE_O2L1_Gauge_Comparison}
\end{center}
\end{table}
The similarities between the gauge content of these two data sets are striking. 
It is tempting to assume that ignoring ST SUSYs when determining uniqueness will result in the data sets being identical, since the ST SUSY generator $\vec{S}$ is the only real difference between the data sets.
This is not the case. 
In fact, ignoring ST SUSYs when comparing the intersection of these two sets of models gives the same number of models common to both sets: 8.
This implies that the matter representations are affected by whether $\vec{S}$ is in the set of basis vectors making up a model.
What is likely occurring in the models with $\vec{S}$ is that the sector coming from $\vec{S}$ is contributing non-adjoint representations to the matter states of the model.
In other words, gravitinos are not the only fermion states coming from $\vec{S}$.
In order to begin fully mapping the heterotic landscape the full effect of leaving out the $\vec{S}$ vector should be examined, as its presence may affect the likelihood of models with three chiral generations. 
Other statistics from this data set will now be presented for completeness.

The number of gauge group factors, ST SUSYs, and $U(1)$ factors are plotted in Figure \ref{fig: NAHE_NSV_O2L1_Stats}.
None of the models had any non-Abelian singlets, suggesting that the non-Abelian singlets may be coming from the $\vec{S}$ sector.
The occurrances of GUT groups are charted in Table \ref{tab: NAHE_NSV_O2L1_GUT_Groups}.
\begin{figure}
\begin{center}
\subfloat[][]{
\begin{tikzpicture}[scale=0.75]
\begin{axis} [ybar, ylabel = Number of Distinct Models, xlabel = Number of Gauge Group Factors]
\addplot[draw=black, fill=black]coordinates{
(4,2) (5,20) (6,30) (7,50) (8,44) (9,56) (10,31) (11,24) (12,14) (13,8) (14,3) };
\end{axis}
\end{tikzpicture}
}
\subfloat[][]{
\begin{tikzpicture}[scale=0.75]
\begin{axis} [nodes near coords, ybar, ylabel = Number of Distinct Models, xlabel = Number of ST SUSYs]
\addplot[draw=black, fill=black]coordinates{
(0,281) (1,1) (2, 0) (4, 0) };
\end{axis}
\end{tikzpicture}
}\\
\subfloat[][]{
\begin{tikzpicture}[scale=0.75]
\begin{axis} [ybar, ylabel = Number of Distinct Models, xlabel = Number of $U(1)$ Factors]
\addplot[draw=black, fill=black]coordinates{
(0,73) (1,105) (2,65) (3,31) (4,8) };
\end{axis}
\end{tikzpicture}
}
\caption{Statistics for the NAHE + O2L1 data set without $\vec{S}$.}
\label{fig: NAHE_NSV_O2L1_Stats}
\end{center}
\end{figure}
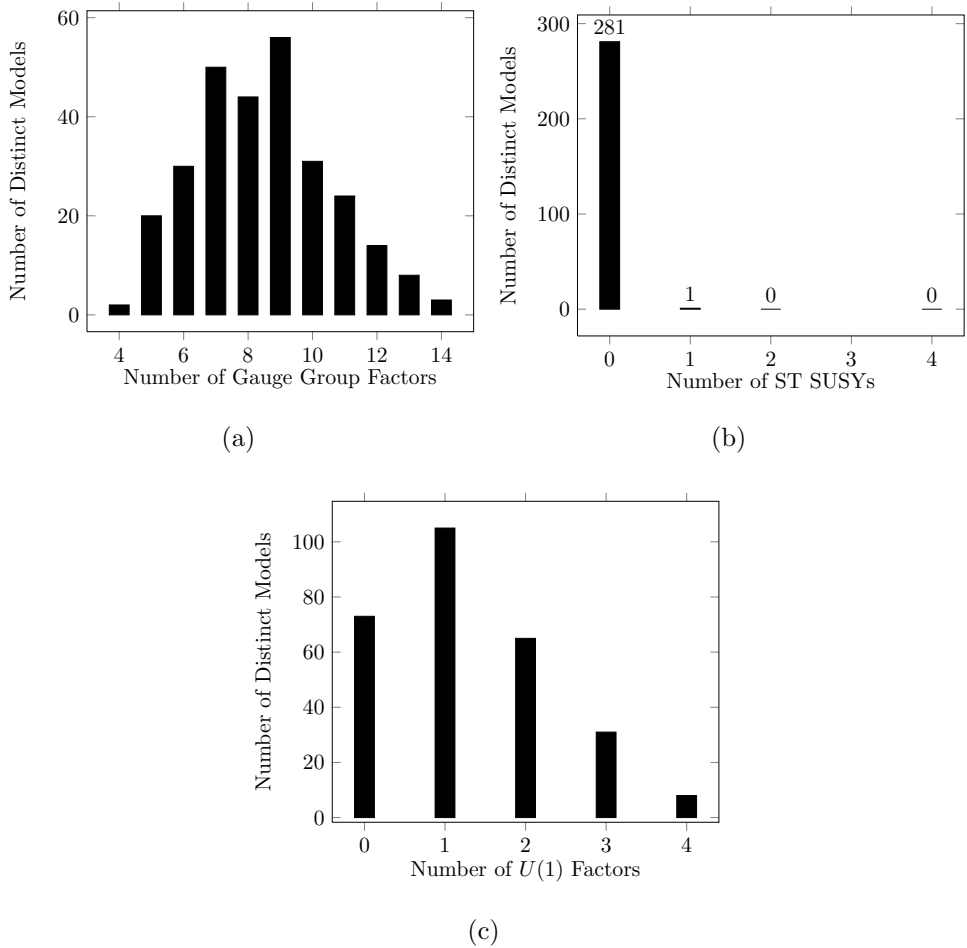
\begin{table}
\caption{The number of unique models containing GUT groups for the NAHE + O2L1 data set without $\vec{S}$.}
\begin{center}
\begin{tabular}{||c|c|c||}
\hline \hline
GUT Group & Number of Unique Models & \% of Unique Models \\
\hline \hline
$E_6$ &1&0.3546\%\\
\hline
$SO(10)$ &100&35.46\%\\
\hline
$SU(5)\otimes U(1)$&0&0\%\\
\hline
$SU(4)\otimes SU(2)\otimes SU(2)$&156&55.32\%\\
\hline
$SU(3)\otimes SU(2)\otimes SU(2)$&0&0\%\\
\hline
$SU(3)\otimes SU(2)\otimes U(1)$&0&0\%\\
\hline \hline
\end{tabular}
\label{tab: NAHE_NSV_O2L1_GUT_Groups}
\end{center}
\end{table}
\section{Statistics for Order-3 Layer-1}\label{sec: NAHE_O3L1_Statistics}
The next set of statistics to be reported are for extensions to the NAHE set with a single basis vector of right moving order-3.
The order-3 basis vectors added are fermion sectors; the left movers are order-2.
Since the orders of the left and right movers are not the same, the total order of the basis vector extensions is 6.
The difference between order-3 basis vectors of this type and true order-6 basis vectors is that all six possibilities for the phases will appear in an order-6 right mover, while only three phases appear in an order-3 right mover.
The coefficients generating the sectors from these basis vectors still range from 0 to 5, however.
This has interesting effects on the fermion spectrum, which have been noted in \cite{Renner:2011}.
Statistics will be presented for models of this type both with and without $\vec{S}$.
\subsection{With $\vec{S}$}
The presence of $\vec{S}$ in the NAHE set for this search causes any order-3 basis vector with the same left mover as $\vec{S}$ to be inconsistent.
This is due to the $\mathds{Z}_2^L || \mathds{Z}_3^R$ symmetries of the left and right mover.
Adding $\vec{S}$ to three times the basis vector extension results in a second $\vec{0}$ sector, which means the set of basis vectors is not linearly independent.
Moreover, since the right mover does not have any periodic phases, there can be no rank-cutting in these models. 
Non-simply laced gauge groups and higher level Ka$\check{c}$-Moody algebras are not present.

\begin{figure}
\begin{center}
\begin{tikzpicture}
\begin{axis} [ybar, ylabel = Number of Distinct Models, xlabel = Number of Gauge Group Factors]
\addplot[draw=black, fill=black]coordinates{
(4,8) (5,98) (6,127) (7,209) (8,267) (9,321) (10,553) (11,473) (12,477) (13,251) (14,200) (15,32) (16,20) };
\end{axis}
\end{tikzpicture}
\caption{The number of gauge group factors per model in the NAHE + O3L1 data set.}
\label{fig: NAHE_O3L1_Gauge_Group_Factors}
\end{center}
\end{figure}
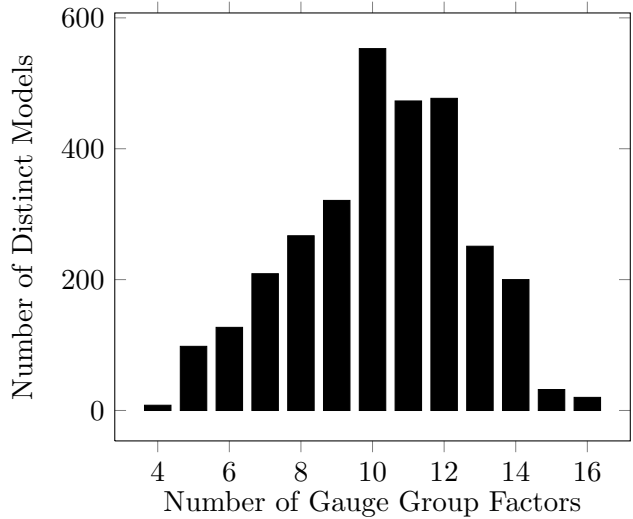

Despite having no linearly independent basis vectors with the same left movers as $\vec{S}$, this data set contains quite a bit more distinct models.
There were 373,152 models in this set, but only 3,036 were unique.
This relatively (compared to the order-2 extension) high number of unique models suggests that the periodic/anti-periodic phases of the order-2 models have a redundancy not present in models of this type.
Based on the double counting of the order-2 models without rank-cuts, the estimated systematic uncertainty for the order-3 statistics is 10\%.
The gauge content of these models is presented in Table \ref{tab: NAHE_O3L1_Gauge_Groups}.
\begin{table}
\caption{The gauge group content of the NAHE + O3L1 data set.}
\begin{center}
\begin{tabular}{||c|c|c||}
\hline \hline
Gauge Group & Number of Unique Models & \% of Unique Models \\
\hline \hline
$SU(2)$&2587&85.21\%\\
\hline
$SU(3)$&923&30.4\%\\
\hline
$SU(4)$&2241&73.81\%\\
\hline
$SU(5)$&543&17.89\%\\
\hline
$SU(6)$&735&24.21\%\\
\hline
$SU(7)$&215&7.082\%\\
\hline
$SU(8)$&460&15.15\%\\
\hline
$SU(9)$&76&2.503\%\\
\hline
$SU(10)$&76&2.503\%\\
\hline
$SU(11)$&17&0.5599\%\\
\hline
$SU(12)$&41&1.35\%\\
\hline
$SU(13)$&3&0.09881\%\\
\hline
$SU(14)$&5&0.1647\%\\
\hline
$SO(8)$&860&28.33\%\\
\hline
$SO(10)$&659&21.71\%\\
\hline
$SO(12)$&400&13.18\%\\
\hline
$SO(14)$&372&12.25\%\\
\hline
$SO(16)$&260&8.564\%\\
\hline
$SO(18)$&11&0.3623\%\\
\hline
$SO(20)$&33&1.087\%\\
\hline
$SO(22)$&5&0.1647\%\\
\hline
$SO(24)$&15&0.4941\%\\
\hline
$SO(26)$&3&0.09881\%\\
\hline
$E_6$&193&6.357\%\\
\hline
$E_7$&147&4.842\%\\
\hline
$E_8$&80&2.635\%\\
\hline
$U(1)$&2955&97.33\%\\
\hline \hline
\end{tabular}
\label{tab: NAHE_O3L1_Gauge_Groups}
\end{center}
\end{table}
The most noticeable difference between Table \ref{tab: NAHE_O3L1_Gauge_Groups} and Table \ref{tab: NAHE_O2L1_Gauge_Groups} is the number of $SU(n+1)$-type gauge groups.
The GSO projections of the new sector break the untwisted ($\vec{0}$) sector from $SO(44)$ to smaller $SO(2n)$-type groups.
Phases that are neither periodic nor anti-periodic transform these groups from $SO(2n)$ to $SU(n)\otimes U(1)$.
Since all of the phases in this set fall into that category, the $SO(2n)$-type groups appear when the states from the added sector are projected out by certain $k_{ij}$ matrix choices.
The $SU(n+1)$-type groups are created when the contributions from the added sector are left in the model.
The number of gauge group factors per model are plotted in Figure \ref{fig: NAHE_O3L1_Gauge_Group_Factors}.
As with Figure \ref{fig: NAHE_O2L1_Gauge_Group_Factors}, there are still very few models which have less than five gauge group factors. 
There are also peaks at 10, 11 and 12 gauge group factors, as opposed to just a single peak at 9 in Figure \ref{fig: NAHE_O2L1_Gauge_Group_Factors}.

\begin{figure}
\begin{center}
\begin{tikzpicture}
\begin{axis} [nodes near coords, enlargelimits=0.15, ybar, ylabel = Number of Distinct Models, xlabel = Number of ST SUSYs]
\addplot[draw=black, fill=black]coordinates{
(0,1445) (1,1305) (2,286) (4, 0) };
\end{axis}
\end{tikzpicture}
\end{center}
\caption{The ST SUSYs for the NAHE + O3L1 data set.}
\label{fig: NAHE_O3L1_ST_SUSYs}
\end{figure}
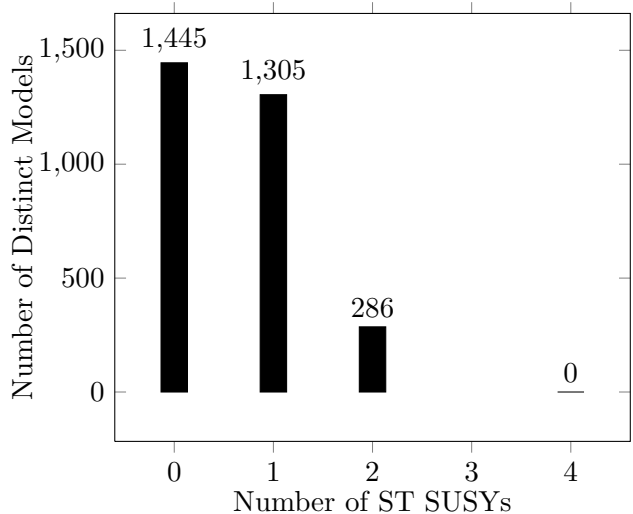

Relevant GUT groups and number of unique models containing those groups are presented in Table \ref{tab: NAHE_O3L1_GUT_Groups}.
The number of ST SUSYs for the order-3 data set is plotted in Figure \ref{fig: NAHE_O3L1_ST_SUSYs}.
Notice that a statistically significant number of models have enhanced ST SUSY.
This trend is expected, as every odd-ordered right mover with a massless fermion left mover will produce an additional gravitino generating sector in the model.
\begin{table}
\caption{The number of unique models containing GUT groups for the NAHE + O3L1 data set.}
\begin{center}
\begin{tabular}{||c|c|c||}
\hline \hline
GUT Group & Number of Unique Models & \% of Unique Models \\
\hline \hline
$E_6$ &193&6.36\%\\
\hline
$SO(10)$&659&21.71\%\\
\hline
$SU(5)\otimes U(1)$&543&17.89\%\\
\hline
$SU(4)\otimes SU(2)\otimes SU(2)$&1648&54.28\%\\
\hline
$SU(3)\otimes SU(2)\otimes SU(2)$&628&20.69\%\\
\hline
$SU(3)\otimes SU(2)\otimes U(1)$&775&25.53\%\\
\hline \hline
\end{tabular}
\label{tab: NAHE_O3L1_GUT_Groups}
\end{center}
\end{table}

\begin{figure}
\begin{center}
\begin{tikzpicture}
\begin{axis} [ybar, ylabel = Number of Distinct Models, xlabel = Number of $U(1)$ Factors]
\addplot[draw=black, fill=black]coordinates{
(0,81) (1,304) (2,518) (3,365) (4,751) (5,586) (6,184) (7,199) (8,36) (9,12) };
\end{axis}
\end{tikzpicture}
\caption{The number of $U(1)$ factors for the NAHE + O3L1 data set.}
\label{fig: NAHE_O3L1_U1_Factors}
\end{center}
\end{figure}
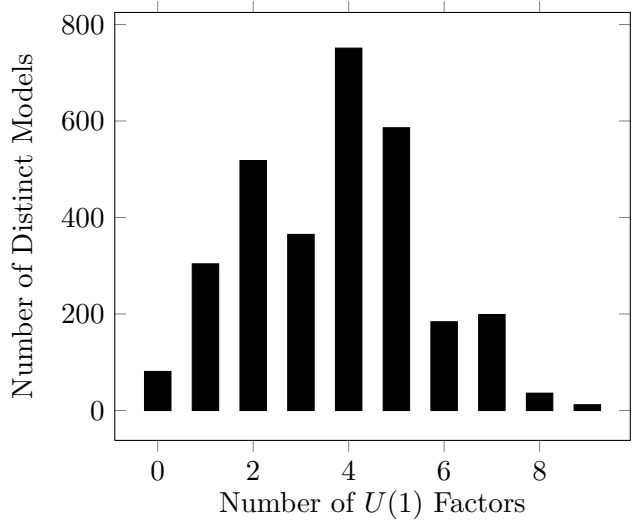

\begin{figure}
\begin{center}
\begin{tikzpicture}
\begin{axis} [ybar, ylabel = Number of Distinct Models,  xlabel = Number of NA Singlets]
\addplot[draw=black, fill=black]coordinates{
(0,1220)(1,18)(2,326)(3,42)(4,577)(5,2)(6,176)(7,12)(8,309)(9,2)(10,46)(11,1)(12,105)(13,7)(14,22)(15,2)(16,65)(17,2)(18,13)(19,1)(20,19)(21,1)(22,10)(24,20)(26,7)(28,3)(30,4)(32,8)(34,1)(36,5)(38,2)(40,3)(42,2)(48,2)(50,1)};
\end{axis}
\end{tikzpicture}
\caption{The number of non-Abelian singlets for the NAHE + O3L1 data set.}
\label{fig: NAHE_O3L1_NA_Singlets}
\end{center}
\end{figure}
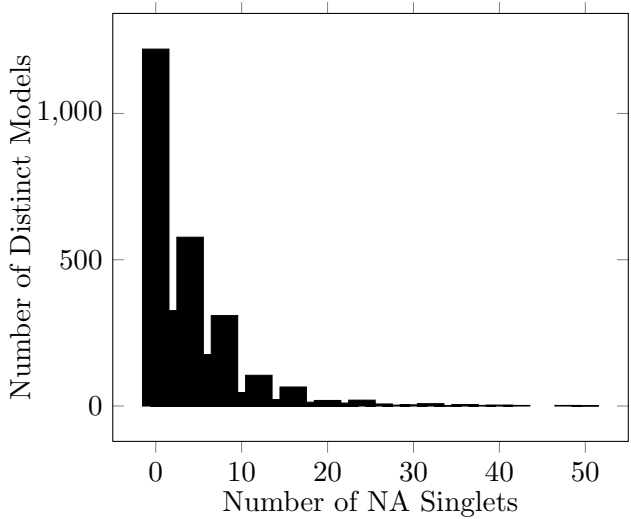

Table \ref{tab: NAHE_O3L1_Gauge_Groups} also makes apparent the number of models containing $U(1)$ gauge groups. 
The number of $U(1)$'s per model are plotted in Figure \ref{fig: NAHE_O3L1_U1_Factors}.
It is clear that most models have multiple $U(1)$ factors, and that there are more $U(1)$ factors per model for this data set than the O2L1 data set.
The number of non-Abelian singlets are plotted in Figure \ref{fig: NAHE_O3L1_NA_Singlets}.
As with the order-2 extensions, the distribution of non-Abelian singlets drops off sharply after 0, indicating the NAHE-base single-layer models do not have a tendency to produce many non-Abelian singlets.
\subsection{Without $\vec{S}$}
Removing $\vec{S}$ from the NAHE set for the order-2 layer-1 extensions had interesting consequences on the available matter sectors.
For the order-3 layer-1 extensions the effect is expected to be more drastic, as linear independence prevented any models with the same left mover as $\vec{S}$ to exist in a model.
There should be less of an impact on the ST SUSY, however, since order-3 basis vectors with massless left movers always produce their own gravitino generating sector.
The lower number of possibilities for $k_{ij}$ values will also have an effect on the number of models in the set.

There are 447 unique models in this set out of 870,688 consistent models, and of those 146 models also belong to the data set with $\vec{S}$. 
Based on the number of duplicates in the O2L1 data set (without $\vec{S}$), the estimated systematic uncertainty for these statistics is 10\%.
There is significantly more overlap between the two sets, yet there are also significantly fewer unique models. 
The gauge group content of the NAHE+O3L1 without $\vec{S}$ is presented in Table \ref{tab: NAHE_NSV_O3L1_Gauge_Groups}.
\begin{table}
\caption{The gauge group content of the NAHE + O3L1 data set without $\vec{S}$.}
\begin{center}
\begin{tabular}{||c|c|c||}
\hline \hline
Gauge Group & Number of Unique Models & \% of Unique Models \\
\hline \hline
$SU(2)$&368&82.33\%\\
\hline
$SU(3)$&128&28.64\%\\
\hline
$SU(4)$&313&70.02\%\\
\hline
$SU(5)$&70&15.66\%\\
\hline
$SU(6)$&96&21.48\%\\
\hline
$SU(7)$&26&5.817\%\\
\hline
$SU(8)$&71&15.88\%\\
\hline
$SU(9)$&15&3.356\%\\
\hline
$SU(10)$&14&3.132\%\\
\hline
$SU(11)$&3&0.6711\%\\
\hline
$SU(12)$&8&1.79\%\\
\hline
$SU(13)$&1&0.2237\%\\
\hline
$SU(14)$&1&0.2237\%\\
\hline
$SO(8)$&124&27.74\%\\
\hline
$SO(10)$&97&21.7\%\\
\hline
$SO(12)$&53&11.86\%\\
\hline
$SO(14)$&49&10.96\%\\
\hline
$SO(16)$&42&9.396\%\\
\hline
$SO(18)$&3&0.6711\%\\
\hline
$SO(20)$&6&1.342\%\\
\hline
$SO(22)$&1&0.2237\%\\
\hline
$SO(24)$&2&0.4474\%\\
\hline
$SO(26)$&1&0.2237\%\\
\hline
$E_6$&37&8.277\%\\
\hline
$E_7$&32&7.159\%\\
\hline
$E_8$&16&3.579\%\\
\hline
$U(1)$&430&96.2\%\\
\hline \hline
\end{tabular}
\label{tab: NAHE_NSV_O3L1_Gauge_Groups}
\end{center}
\end{table}
A brief comparison between Tables \ref{tab: NAHE_NSV_O3L1_Gauge_Groups} and \ref{tab: NAHE_O3L1_Gauge_Groups} makes it clear the presence of $\vec{S}$ did not significantly affect the gauge groups, as was the case with the order-2 extensions.
For completeness the occurances of the GUT groups and other relevant statistics for this data set will be presented.
The occurrances of the GUT groups in this data set are tabulated in Table \ref{tab: NAHE_NSV_O3L1_GUT_Groups}.
The number of gauge group factors, $U(1)$ factors, ST SUSYs, and non-Abelian singlets are plotted in Figure \ref{fig: NAHE_NSV_O3L1_Statistics}. 
\begin{table}[h]
\caption{The occurrances of the GUT groups for the NAHE + O3L1 data set without $\vec{S}$.}
\begin{center}
\begin{tabular}{||c|c|c||}
\hline \hline
Gauge Group & Number of Unique Models & \% of Unique Models \\
\hline \hline
$E_6$&37&8.277\%\\
\hline
$SO(10)$&97&21.7\%\\
\hline
$SU(5)\otimes U(1)$&70&15.66\%\\
\hline
$SU(4)\otimes SU(2)\otimes SU(2)$&220&49.21\%\\
\hline
$SU(3)\otimes SU(2)\otimes SU(2)$&81&18.12\%\\
\hline
$SU(3)\otimes SU(2)\otimes U(1)$&100&22.37\%\\
\hline \hline
\end{tabular}
\label{tab: NAHE_NSV_O3L1_GUT_Groups}
\end{center}
\end{table}
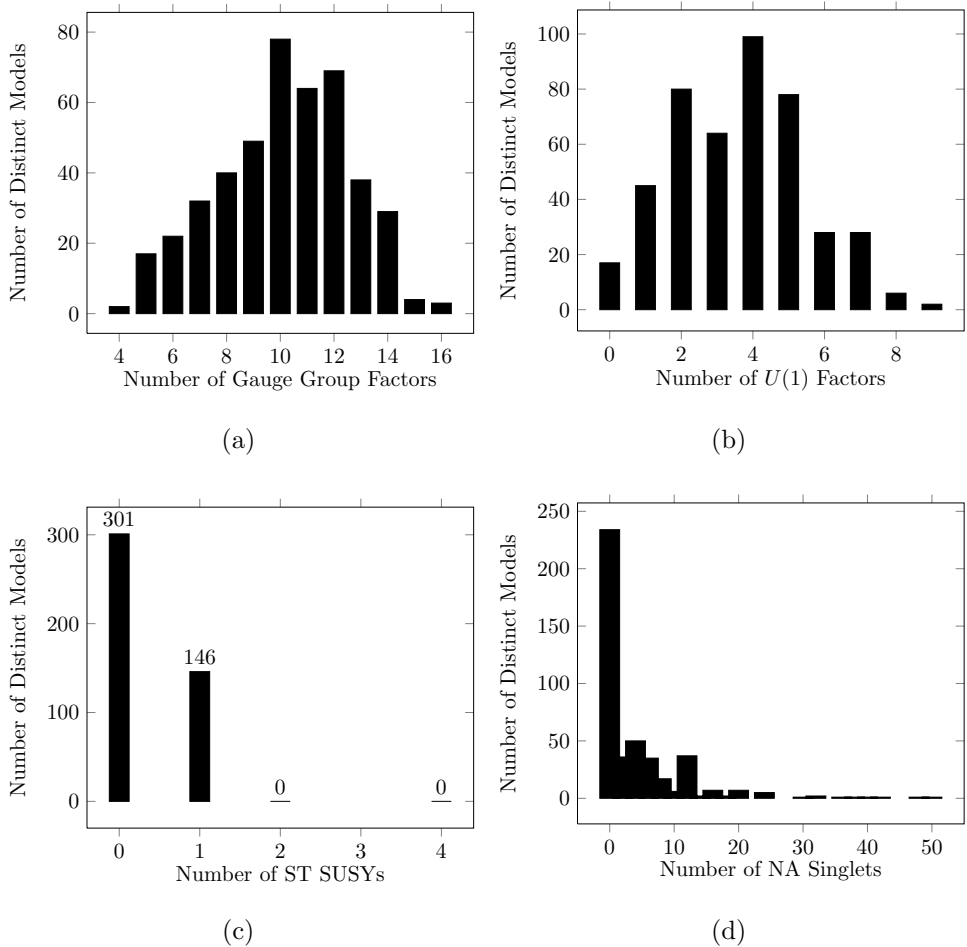
\begin{figure}
\begin{center}
\subfloat[][]{
\begin{tikzpicture}[scale=0.75]
\begin{axis} [ybar, ylabel = Number of Distinct Models, xlabel = Number of Gauge Group Factors]
\addplot[draw=black, fill=black]coordinates{
(4,2) (5,17) (6,22) (7,32) (8,40) (9,49) (10,78) (11,64) (12,69) (13,38) (14,29) (15,4) (16,3) };
\end{axis}
\end{tikzpicture}
}
\subfloat[][]{
\begin{tikzpicture}[scale=0.75]
\begin{axis} [ybar, ylabel = Number of Distinct Models, xlabel = Number of $U(1)$ Factors]
\addplot[draw=black, fill=black]coordinates{
(0,17) (1,45) (2,80) (3,64) (4,99) (5,78) (6,28) (7,28) (8,6) (9,2) };
\end{axis}
\end{tikzpicture}
}\\
\subfloat[][]{
\begin{tikzpicture}[scale=0.75]
\begin{axis} [nodes near coords, ybar, ylabel = Number of Distinct Models, xlabel = Number of ST SUSYs]
\addplot[draw=black, fill=black]coordinates{
(0,301) (1,146) (2, 0) (4, 0) };
\end{axis}
\end{tikzpicture}
}
\subfloat[][]{
\begin{tikzpicture}[scale=0.75]
\begin{axis} [ybar, ylabel = Number of Distinct Models,  xlabel = Number of NA Singlets]
\addplot[draw=black, fill=black]coordinates{
(0,234)(2,36)(4,50)(6,35)(8,17)(10,6)(12,37)(14,2)(16,7)(18,2)(20,7)(24,5)(30,1)(32,2)(36,1)(38,1)(40,1)(42,1)(48,1)(50,1)};
\end{axis}
\end{tikzpicture}
}\\
\caption{Statistics for the NAHE + O3L1 data set without $\vec{S}$.}
\label{fig: NAHE_NSV_O3L1_Statistics} 
\end{center}
\end{figure}
\section{Models With GUT Groups}\label{sec: NAHE_Models_With_GUT_Groups}
The next several sections will outline statistics on models containing GUT groups from the NAHE + O2L1 and NAHE + O3L1 data sets (with $\vec{S}$).
The GUT groups to be examined are $E_6$, $SO(10)$, $SU(5)\otimes U(1)$, $SU(4)\otimes SU(2) \otimes SU(2)$ (Pati-Salam), $SU(3)\otimes SU(2)\otimes SU(2)$ (Left-Right Symmetric), and $SU(3)\otimes SU(2)\otimes U(1)$ (MSSM).
In addition to the spread of statistics presented in sections \ref{sec: NAHE_O2L1_Statistics} and \ref{sec: NAHE_O3L1_Statistics}, the number of chiral fermion generations will be counted, along with any exotics which carry observable sector charge.
These statistics will be gathered for all possible observable sector configurations.
Thus, if there is more than one copy of any GUT group in a model (as is often the case), all possible choices for each group forming the observable sector will be examined.

The definition of a chiral matter generation will be presented with each group for clarity, and only the ``net" chiral generations will be counted.
That is, if there are an equal number of barred and unbarred generations, the model will have no net chiral generations.
Any ``net" generations which cannot be paired must remain massless until the Higgs boson gains a VEV at the TeV scale.

The statistics gathered on the chiral generations here are not enough to qualify a model or set of models as being ``realistic." 
This study looks only to examine the basic components that sometimes lead to realistic models.
Actually determining whether or not a model is realistic requires detailed analysis of the $U(1)$ charges, finding the superpotential, and finding the D- and F-flat directions.
Progress is being made to automate the above steps and integrate the deeper phenomenological components of WCFFHS model building into the FF Framework.
For the present analysis, however, discussion of these aspects of WCFFHS phenomenology is omitted.
\subsection{$E_6$ Models}
Each SM generation of fermions fits into a 27 dimensional representation of $E_6$, so the number of net chiral matter generations is given by 
\begin{equation}
|N_{27} - N_{\overline{27}}|.\label{eqn: E6_Net_Generations}
\end{equation}
Additionally, for large GUT groups, states which transform under the "hidden sector" can be treated as being multiple copies of an observable generation.
The hidden sector groups can be broken somewhat easily by adding basis vectors, so for certain GUT groups the dimension of the hidden sector charge is treated as the number of duplicate observable generations.

\begin{figure}
\begin{center}
\subfloat[][With Hidden Sector Duplicates]{
\begin{tikzpicture}[scale=0.75]
\begin{axis} [ybar, ylabel = Number of Distinct Observable Sectors,  xlabel = Number of Chiral Matter Generations]
\addplot[draw=black, fill=black]coordinates{
(0,192)(2,2)(4,2)(8,4)(16,2)(24,1)};
\end{axis}
\end{tikzpicture}
}
\subfloat[][Without Hidden Sector Duplicates]{
\begin{tikzpicture}[scale=0.75]
\begin{axis} [ybar, ylabel = Number of Distinct Observable Sectors,  xlabel = Number of Chiral Matter Generations]
\addplot[draw=black, fill=black]coordinates{
(0,198)(2,2)(4,2)(8,1)};
\end{axis}
\end{tikzpicture}
}\\
\subfloat[][]{
\begin{tikzpicture}[scale=0.75]
\begin{axis} [ybar, ylabel = Number of Distinct Models,  xlabel = Number of Charged Exotics]
\addplot[draw=black, fill=black]coordinates{
(0,168)(2,6)(4,19)(6,2)(8,7)(12,1)};
\end{axis}
\end{tikzpicture}
}
\caption{Statistics related to the chiral matter generations for $E_6$ models in the NAHE + O3L1 data set.}
\label{fig: E6_NAHE_O3L1_OS_Statistics}
\end{center}
\end{figure}
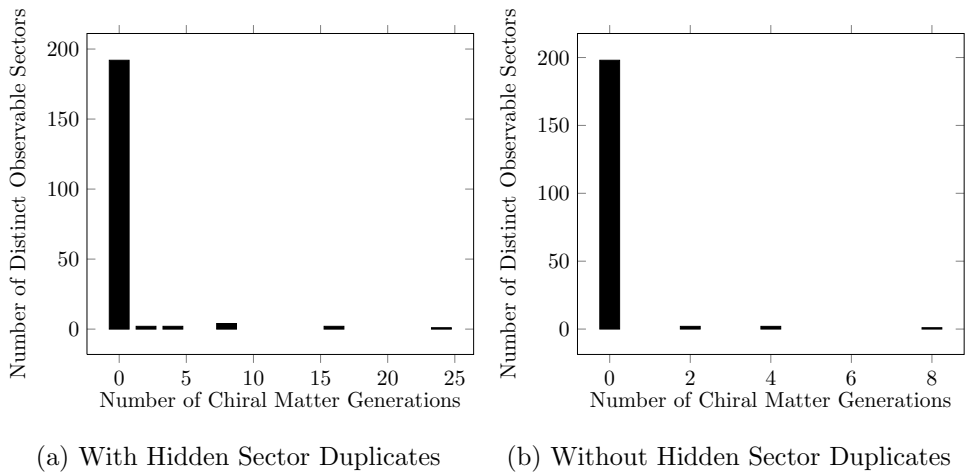
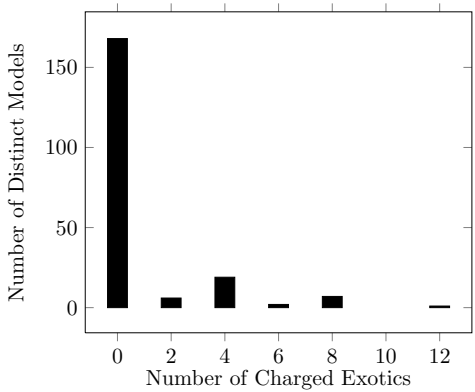

Statistics will now be presented for $E_6$ models coming from single layer extensions to the NAHE set.
There was only one model with an $E_6$ group in the NAHE + O2L1 data set, but there were 193 models with $E_6$ in the NAHE + O3L1 data set.
The number of net chiral fermion generations with and without hidden sector duplicates is plotted in Figure \ref{fig: E6_NAHE_O3L1_OS_Statistics} along with the number of observable sector charged exotics.
The hidden sector gauge group content of these models is presented in Table \ref{tab: E6_NAHE_O3L1_Gauge_Groups}.
\begin{table}
\caption{The hidden sector gauge group content of models containing $E_6$ within the NAHE + O3L1 data set.}
\begin{center}
\begin{tabular}{||c|c|c||}
\hline \hline
Gauge Group & Number of Unique Models & \% of Unique Models \\
\hline \hline
$SU(2)$&165&85.49\%\\
\hline
$SU(3)$&54&27.98\%\\
\hline
$SU(4)$&113&58.55\%\\
\hline
$SU(5)$&39&20.21\%\\
\hline
$SU(6)$&43&22.28\%\\
\hline
$SU(7)$&5&2.591\%\\
\hline
$SU(8)$&10&5.181\%\\
\hline
$SO(8)$&35&18.13\%\\
\hline
$SO(10)$&41&21.24\%\\
\hline
$SO(12)$&12&6.218\%\\
\hline
$SO(16)$&19&9.845\%\\
\hline
$E_8$&11&5.699\%\\
\hline
$U(1)$&193&100\%\\
\hline \hline
\end{tabular}
\label{tab: E6_NAHE_O3L1_Gauge_Groups}
\end{center}
\end{table}
Of note in this table is the presence of $SU(5)$ and $SU(3)$ hidden sector gauge groups in large quantities.
This result stands in contrast to those found in the order-2 search performed in \cite{Dienes:2006}, and is likely due to the extension being an order-3 basis vector.
Similar conclusions were also found in \cite{Moore:2011}.
Order-3 basis vector extensions have very different properties from order-2 and order-4 basis vectors.
Figures \ref{fig: E6_NAHE_O3L1_OS_Statistics} shows that the distribution of net fermion generations tends to be at zero; either no $27$'s are produced, or every $27$ is accompanied by a $\overline{27}$ for those models.
For models with more than zero chiral fermion generations, the number of generations per model is even both with and without hidden sector duplicates.
It is also apparent from Figure \ref{fig: E6_NAHE_O3L1_OS_Statistics} that the $E_6$ charged fermions do not couple to the hidden sector in most of these models.
These examinations make it clear that, though most models do not contain exotic states, there are never the correct number of chiral fermion generations to produce a realistic model.

\begin{figure}
\begin{center}
\subfloat[][]{
\begin{tikzpicture}[scale=0.75]
\begin{axis} [ybar, ylabel = Number of Distinct Models, xlabel = Number of Gauge Group Factors]
\addplot[draw=black, fill=black]coordinates{
(5,7) (6,5) (7,26) (8,12) (9,47) (10,32) (11,39) (12,17) (13,8) };
\end{axis}
\end{tikzpicture}
}
\subfloat[][]{
\begin{tikzpicture}[scale=0.75]
\begin{axis} [ybar, ylabel = Number of Distinct Models, xlabel = Number of $U(1)$ Factors]
\addplot[draw=black, fill=black]coordinates{
(1,5) (2,52) (3,69) (4,39) (5,10) (6,18) };
\end{axis}
\end{tikzpicture}
}\\

\subfloat[][]{
\begin{tikzpicture}[scale=0.75]
\begin{axis} [nodes near coords, ybar, ylabel = Number of Distinct Models, xlabel = Number of ST SUSYs]
\addplot[draw=black, fill=black]coordinates{
(0,81) (1,48) (2,64) (4, 0) };
\end{axis}
\end{tikzpicture}
}
\subfloat[][]{
\begin{tikzpicture}[scale=0.75]
\begin{axis} [ybar, ylabel = Number of Distinct Models,  xlabel = Number of NA Singlets]
\addplot[draw=black, fill=black]coordinates{
(0,100)(2,18)(4,33)(6,8)(8,19)(12,9)(16,6)};
\end{axis}
\end{tikzpicture}
}
\caption{Statistics for the models containing $E_6$ in the NAHE + O3L1 data set.}
\label{fig: E6_NAHE_O3L1_Statistics}
\end{center}
\end{figure}
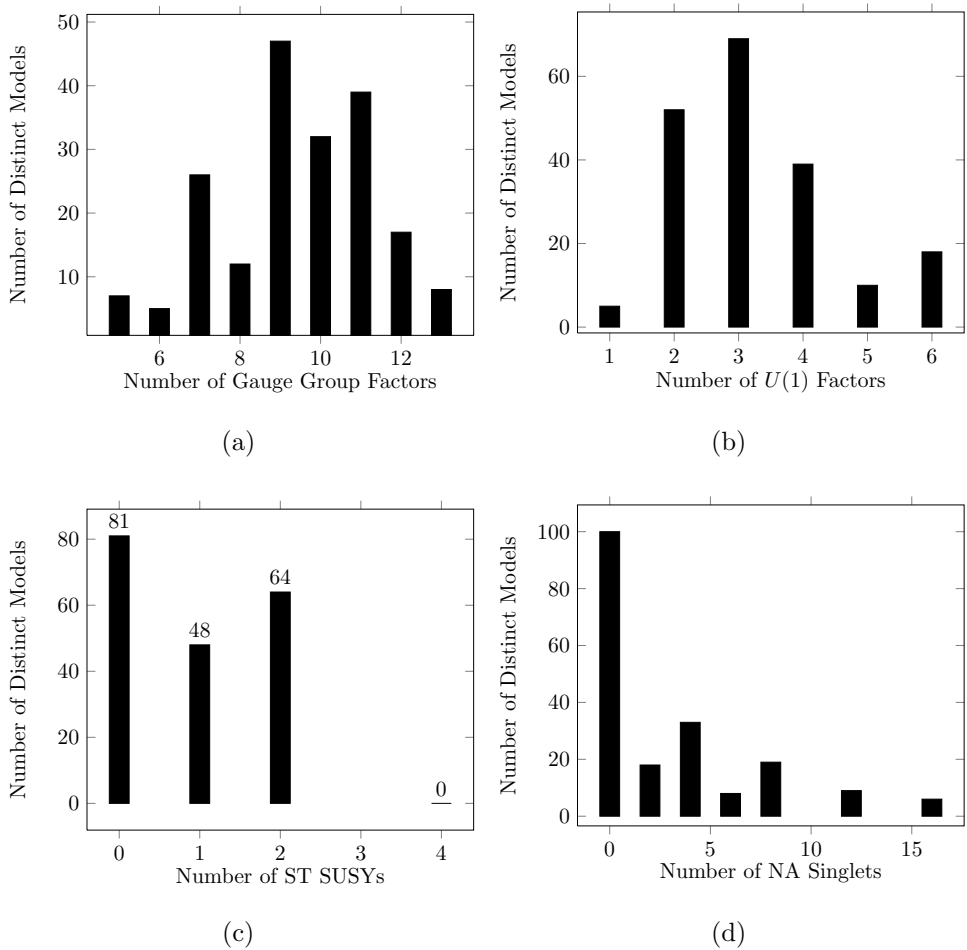

For completeness the rest of the statistics for this data set will be presented. The number of gauge group factors per model, the number of $U(1)$ factors, the number of ST SUSYs, and the number of non-Abelian singlets are presented in Figure \ref{fig: E6_NAHE_O3L1_Statistics}.
\subsection{$SO(10)$ Models}
In the models containing $SO(10)$, the chiral fermion generations are defined to be in the 16 dimensional representations.
Thus, the number of net chiral fermion generations is given by 
\begin{equation}
|N_{16} - N_{\overline{16}}|.\label{eqn: SO10_Net_Generations}
\end{equation}
The NAHE + O2L1 data set has 160 models with $SO(10)$, while the NAHE + O3L1 data set has 659 models.
The number of chiral fermion generations with hidden sector charges is plotted in Figure \ref{fig: SO10_NAHE_Generations} for the NAHE + O2L1 and NAHE + O3L1 data sets.
The number of chiral fermion generations without hidden sector charges is also plotted in Figure \ref{fig: SO10_NAHE_Generations} for the order-3 $SO(10)$ models. 
There were no order-2 $SO(10)$ models with more than zero chiral fermion generations.
The hidden sector gauge groups for the NAHE + O2L1 $SO(10)$ models are presented in Table \ref{tab: SO10_NAHE_O2L1_Gauge_Groups}. The NAHE + O3L1 $SO(10)$ model hidden sector gauge groups are presented in Table \ref{tab: SO10_NAHE_O3L1_Gauge_Groups}.
The number of observable exotic states for these models is plotted in Figure \ref{fig: SO10_NAHE_Exotics} for the order-2 and order-3 models.
\begin{figure}
\begin{center}
\subfloat[][NAHE + O2L1 with Hidden Sector Duplicates]{
\begin{tikzpicture}[scale=0.75]
\begin{axis} [ybar, ylabel = Number of Distinct Observable Sectors,  xlabel = Number of Chiral Matter Generations]
\addplot[draw=black, fill=black]coordinates{
(0,16)(12,4)(16,12)(24,104)(32,12)(36,4)(48,10)};
\end{axis}
\end{tikzpicture}
}
\subfloat[][NAHE + O3L1 with Hidden Sector Duplicates]{
\begin{tikzpicture}[scale=0.75]
\begin{axis} [ybar, ylabel = Number of Distinct Observable Sectors,  xlabel = Number of Chiral Matter Generations]
\addplot[draw=black, fill=black]coordinates{
(0,433)(2,16)(4,46)(6,14)(8,80)(10,10)(12,30)(14,4)(16,55)(20,10)(22,2)(24,18)(26,2)(32,8)(36,2)(48,1)};
\end{axis}
\end{tikzpicture}
}\\
\subfloat[][NAHE + O3L1 without Hidden Sector Duplicates]{
\begin{tikzpicture}[scale=0.75]
\begin{axis} [ybar, ylabel = Number of Distinct Observable Sectors,  xlabel = Number of Chiral Matter Generations]
\addplot[draw=black, fill=black]coordinates{
(0,598)(2,36)(4,69)(6,12)(8,15)(12,1)};
\end{axis}
\end{tikzpicture}
}
\caption{Statistics for the chiral matter generations of the $SO(10)$ models in the NAHE + O2L1 and NAHE + O3L1 data sets.}
\label{fig: SO10_NAHE_Generations}
\end{center}
\end{figure}
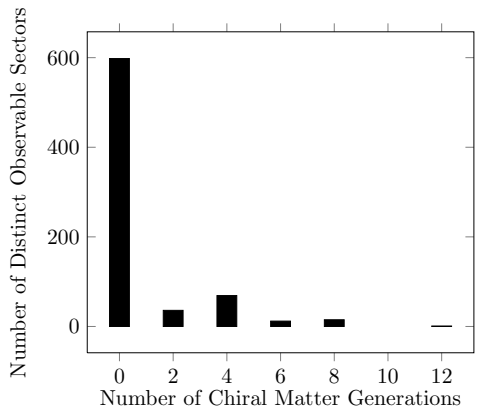
\begin{table}
\caption{Hidden sector gauge groups for $SO(10)$ models in the NAHE + O2L1 data set.}
\begin{center}
\begin{tabular}{||c|c|c||}
\hline \hline
Gauge Group & Number of Unique Models & \% of Unique Models \\
\hline \hline
$SU(2)$&121&75.62\%\\
\hline
$SU(2)^{(2)}$&25&15.62\%\\
\hline
$SU(4)$&111&69.38\%\\
\hline
$SO(5)$&59&36.88\%\\
\hline
$SO(8)$&1&0.625\%\\
\hline
$SO(14)$&1&0.625\%\\
\hline
$SO(16)$&56&35\%\\
\hline
$SO(20)$&1&0.625\%\\
\hline
$SO(22)$&1&0.625\%\\
\hline
$E_7$&51&31.87\%\\
\hline
$E_8$&51&31.87\%\\
\hline
$U(1)$&102&63.75\%\\
\hline \hline
\end{tabular}
\label{tab: SO10_NAHE_O2L1_Gauge_Groups}
\end{center}
\end{table}
\begin{table}
\caption{Hidden sector gauge groups for $SO(10)$ models in the NAHE + O3L1 data set.}
\begin{center}
\begin{tabular}{||c|c|c||}
\hline \hline
Gauge Group & Number of Unique Models & \% of Unique Models \\
\hline \hline
$SU(2)$&570&86.49\%\\
\hline
$SU(3)$&126&19.12\%\\
\hline
$SU(4)$&456&69.2\%\\
\hline
$SU(5)$&74&11.23\%\\
\hline
$SU(6)$&132&20.03\%\\
\hline
$SU(7)$&20&3.035\%\\
\hline
$SU(8)$&64&9.712\%\\
\hline
$SU(9)$&5&0.7587\%\\
\hline
$SU(10)$&15&2.276\%\\
\hline
$SU(12)$&12&1.821\%\\
\hline
$SO(8)$&105&15.93\%\\
\hline
$SO(12)$&75&11.38\%\\
\hline
$SO(14)$&34&5.159\%\\
\hline
$SO(16)$&69&10.47\%\\
\hline
$SO(20)$&13&1.973\%\\
\hline
$SO(22)$&5&0.7587\%\\
\hline
$E_6$&41&6.222\%\\
\hline
$E_7$&29&4.401\%\\
\hline
$E_8$&20&3.035\%\\
\hline
$U(1)$&604&91.65\%\\
\hline \hline
\end{tabular}
\label{tab: SO10_NAHE_O3L1_Gauge_Groups}
\end{center}
\end{table}
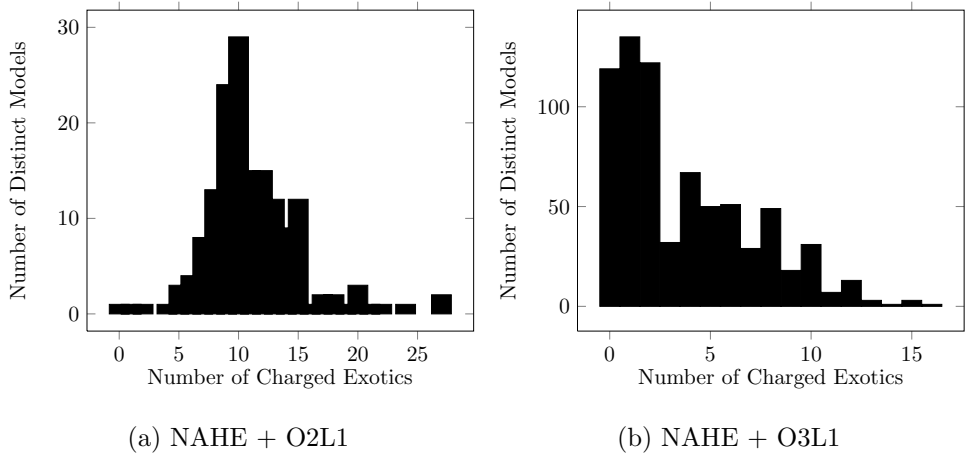
\begin{figure}
\begin{center}
\subfloat[][NAHE + O2L1]{
\begin{tikzpicture}[scale=0.75]
\begin{axis} [ybar, ylabel = Number of Distinct Models,  xlabel = Number of Charged Exotics]
\addplot[draw=black, fill=black]coordinates{
(0,1)(1,1)(2,1)(4,1)(5,3)(6,4)(7,8)(8,13)(9,24)(10,29)(11,15)(12,15)(13,12)(14,9)(15,12)(16,1)(17,2)(18,2)(19,1)(20,3)(21,1)(22,1)(24,1)(27,2)};
\end{axis}
\end{tikzpicture}
}
\subfloat[][NAHE + O3L1]{
\begin{tikzpicture}[scale=0.75]
\begin{axis} [ybar, ylabel = Number of Distinct Models,  xlabel = Number of Charged Exotics]
\addplot[draw=black, fill=black]coordinates{
(0,119)(1,135)(2,122)(3,32)(4,67)(5,50)(6,51)(7,29)(8,49)(9,18)(10,31)(11,7)(12,13)(13,3)(14,1)(15,3)(16,1)};
\end{axis}
\end{tikzpicture}
}
\caption{The number of observable sector charged exotics for $SO(10)$ models in the NAHE + O2L1 and NAHE + O3L1 data sets.}
\label{fig: SO10_NAHE_Exotics}
\end{center}
\end{figure}
As with the $E_6$ data sets, there are no models with three chiral generations.
The number of generations in any of the models presented is either zero or even.
The distribution of charged exotics is more spread out than it was for the $E_6$ models.
This is an artifact of the NAHE set gauge group; the $SO(10)$ states in that set are charged under the three $SU(4)$ gauge groups.
Most observable $SO(10)$ states tend to keep some of those charges under the hidden sector.
The remaining statistics for these models are presented in Figure \ref{fig: SO10_NAHE_O2L1_Statistics} for the order-2 models.
\begin{figure}
\begin{center}
\subfloat[][]{
\begin{tikzpicture}[scale=0.75]
\begin{axis} [ybar, ylabel = Number of Distinct Models, xlabel = Number of Gauge Group Factors]
\addplot[draw=black, fill=black]coordinates{
(4,1) (5,34) (6,20) (7,31) (8,24) (9,28) (10,8) (11,11) (12,3) };
\end{axis}
\end{tikzpicture}
}
\subfloat[][]{
\begin{tikzpicture}[scale=0.75]
\begin{axis} [ybar, ylabel = Number of Distinct Models, xlabel = Number of $U(1)$ Factors]
\addplot[draw=black, fill=black]coordinates{
(0,58) (1,60) (2,27) (3,15) };
\end{axis}
\end{tikzpicture}
}\\
\subfloat[][]{
\begin{tikzpicture}[scale=0.75]
\begin{axis} [nodes near coords, ybar, ylabel = Number of Distinct Models, xlabel = Number of ST SUSYs]
\addplot[draw=black, fill=black]coordinates{
(0,82) (1,77) (2,1) (4, 0) };
\end{axis}
\end{tikzpicture}
}
\subfloat[][]{
\begin{tikzpicture}[scale=0.75]
\begin{axis} [ybar, ylabel = Number of Distinct Models,  xlabel = Number of NA Singlets]
\addplot[draw=black, fill=black]coordinates{
(0,66)(2,19)(4,35)(6,16)(8,20)(12,4)};
\end{axis}
\end{tikzpicture}
}
\caption{Statistics for the $SO(10)$ models in the NAHE + O2L1 data set.}
\label{fig: SO10_NAHE_O2L1_Statistics}
\end{center}
\end{figure}
The remaining statistics for the order-3 models with $SO(10)$ are presented in Figure \ref{fig: SO10_NAHE_O3L1_Statistics}.
\begin{figure}
\begin{center}
\subfloat[][]{
\begin{tikzpicture}[scale=0.75]
\begin{axis} [ybar, ylabel = Number of Distinct Models, xlabel = Number of Gauge Group Factors]
\addplot[draw=black, fill=black]coordinates{
(4,5) (5,48) (6,33) (7,66) (8,73) (9,81) (10,118) (11,82) (12,109) (13,32) (14,12) };
\end{axis}
\end{tikzpicture}
}
\subfloat[][]{
\begin{tikzpicture}[scale=0.75]
\begin{axis} [ybar, ylabel = Number of Distinct Models, xlabel = Number of $U(1)$ Factors]
\addplot[draw=black, fill=black]coordinates{
(0,55) (1,78) (2,111) (3,99) (4,171) (5,87) (6,30) (7,28) };
\end{axis}
\end{tikzpicture}
}\\
\subfloat[][]{
\begin{tikzpicture}[scale=0.75]
\begin{axis} [nodes near coords, ybar, ylabel = Number of Distinct Models, xlabel = Number of ST SUSYs]
\addplot[draw=black, fill=black]coordinates{
(0,318) (1,294) (2,47) (4, 0) };
\end{axis}
\end{tikzpicture}
}
\subfloat[][]{
\begin{tikzpicture}[scale=0.75]
\begin{axis} [ybar, ylabel = Number of Distinct Models,  xlabel = Number of NA Singlets]
\addplot[draw=black, fill=black]coordinates{
(0,301)(1,10)(2,60)(3,4)(4,120)(6,40)(8,67)(9,1)(10,13)(12,17)(13,2)(14,2)(16,11)(17,1)(18,1)(20,6)(24,2)(28,1)};
\end{axis}
\end{tikzpicture}
}
\caption{Statistics for the $SO(10)$ models in the NAHE + O3L1 data set.}
\label{fig: SO10_NAHE_O3L1_Statistics}
\end{center}
\end{figure}
\subsection{$SU(5)\otimes U(1)$ Models}
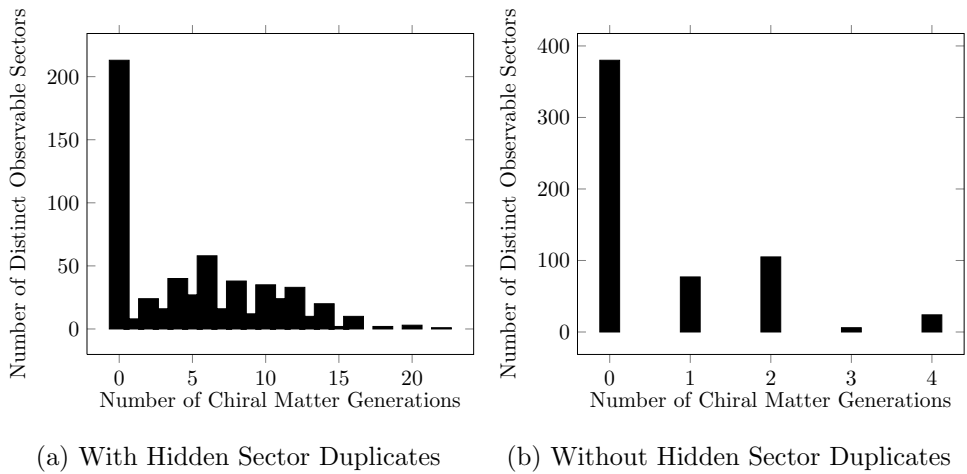
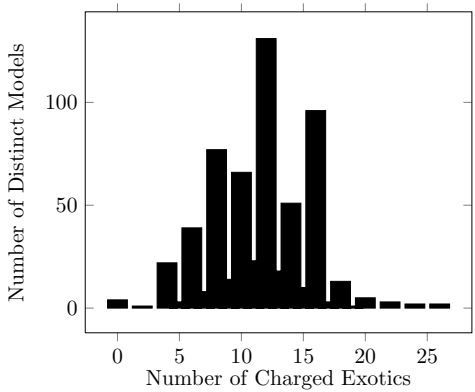
\begin{figure}
\begin{center}
\subfloat[][With Hidden Sector Duplicates]{
\begin{tikzpicture}[scale=0.75]
\begin{axis} [ybar, ylabel = Number of Distinct Observable Sectors,  xlabel = Number of Chiral Matter Generations]
\addplot[draw=black, fill=black]coordinates{
(0,213)(1,8)(2,24)(3,16)(4,40)(5,27)(6,58)(7,16)(8,38)(9,12)(10,35)(11,24)(12,33)(13,10)(14,20)(15,2)(16,10)(18,2)(20,3)(22,1)};
\end{axis}
\end{tikzpicture}
}
\subfloat[][Without Hidden Sector Duplicates]{
\begin{tikzpicture}[scale=0.75]
\begin{axis} [ybar, ylabel = Number of Distinct Observable Sectors,  xlabel = Number of Chiral Matter Generations]
\addplot[draw=black, fill=black]coordinates{
(0,380)(1,77)(2,105)(3,6)(4,24)};
\end{axis}
\end{tikzpicture}
}\\
\subfloat[][]{
\begin{tikzpicture}[scale=0.75]
\begin{axis} [ybar, ylabel = Number of Distinct Models,  xlabel = Number of Charged Exotics]
\addplot[draw=black, fill=black]coordinates{
(0,4)(2,1)(4,22)(5,3)(6,39)(7,8)(8,77)(9,14)(10,66)(11,23)(12,131)(13,18)(14,51)(15,10)(16,96)(17,3)(18,13)(19,1)(20,5)(22,3)(24,2)(26,2)};
\end{axis}
\end{tikzpicture}
}
\caption{Statistics related to observable matter for the $SU(5)\otimes U(1)$ models in the NAHE + O3L1 data set.}
\label{fig: SU5_NAHE_O3L1_OS_Statistics}
\end{center}
\end{figure}

\begin{figure}
\begin{center}
\subfloat[][]{
\begin{tikzpicture}[scale=0.75]
\begin{axis} [ybar, ylabel = Number of Distinct Models, xlabel = Number of Gauge Group Factors]
\addplot[draw=black, fill=black]coordinates{
(9,37) (10,107) (11,164) (12,105) (13,74) (14,48) (15,8) };
\end{axis}
\end{tikzpicture}
}
\subfloat[][]{
\begin{tikzpicture}[scale=0.75]
\begin{axis} [ybar, ylabel = Number of Distinct Models, xlabel = Number of $U(1)$ Factors]
\addplot[draw=black, fill=black]coordinates{
(3,27) (4,188) (5,202) (6,58) (7,64) (9,4) };
\end{axis}
\end{tikzpicture}
}\\
\subfloat[][]{
\begin{tikzpicture}[scale=0.75]
\begin{axis} [nodes near coords, ybar, ylabel = Number of Distinct Models, xlabel = Number of ST SUSYs]
\addplot[draw=black, fill=black]coordinates{
(0,262) (1,243) (2,38) (4, 0) };
\end{axis}
\end{tikzpicture}
}
\subfloat[][]{
\begin{tikzpicture}[scale=0.75]
\begin{axis} [ybar, ylabel = Number of Distinct Models,  xlabel = Number of NA Singlets]
\addplot[draw=black, fill=black]coordinates{
(0,190)(1,18)(2,67)(3,42)(4,73)(5,2)(6,25)(7,12)(8,45)(9,2)(10,11)(11,1)(12,15)(13,7)(14,11)(15,2)(16,3)(17,2)(18,1)(19,1)(21,1)(22,1)(24,3)(26,3)(28,1)(34,1)(40,2)(42,1)};
\end{axis}
\end{tikzpicture}
}
\caption{Statistics for the $SU(5)\otimes U(1)$ models in the NAHE + O3L1 data set.}
\label{fig: SU5_NAHE_O3L1_Statistics}
\end{center}
\end{figure}
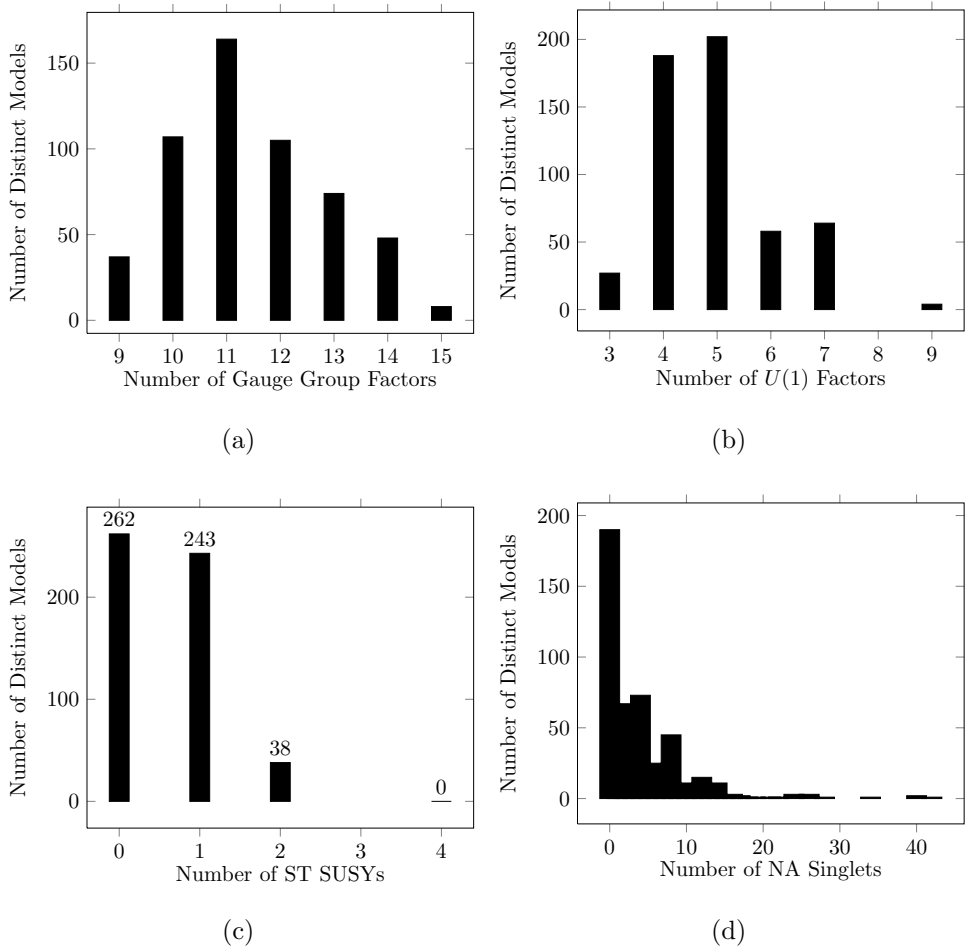

The (flipped) $SU(5)$ GUT group's matter generations are split into multiple representations of the $SU(5)$ group.
An anti-lepton doublet and the up-type quarks are placed in a $\overline{5}$ representation, while the right-handed neutrino, the anti-quark doublet, and the down-type quarks appear in a 10 dimensional representation of $SU(5)$.
Thus, a generation is formed by pairing the 10-reps with the $\overline{5}$-reps. 
The net number of generations for an $SU(5)\otimes U(1)$ model is given by
\begin{equation}
|\text{min}(N_{10},N_{\overline{5}}) -\text{min}(N_{\overline{10}},N_5)|.\label{eqn: SU5_Net_Generations}
\end{equation}
There were no order-2 NAHE based models containing $SU(5)\otimes U(1)$, but there were 543 order-3 models with this GUT group.
The hidden sector gauge groups of those models are presented in Table \ref{tab: SU5_NAHE_O3L1_Gauge_Groups}. 
The number of net chiral generations with and without hidden sector duplicates are plotted in Figure \ref{fig: SU5_NAHE_O3L1_OS_Statistics}, along with the number of exotic states with observable sector charges.
\begin{table}
\caption{The hidden sector gauge groups of the $SU(5)\otimes U(1)$ models in the NAHE + O3L1 data set.}
\begin{center}
\begin{tabular}{||c|c|c||}
\hline \hline
Gauge Group & Number of Unique Models & \% of Unique Models \\
\hline \hline
$SU(2)$&449&82.69\%\\
\hline
$SU(3)$&468&86.19\%\\
\hline
$SU(4)$&284&52.3\%\\
\hline
$SU(6)$&52&9.576\%\\
\hline
$SU(7)$&50&9.208\%\\
\hline
$SU(8)$&52&9.576\%\\
\hline
$SU(9)$&22&4.052\%\\
\hline
$SU(10)$&5&0.9208\%\\
\hline
$SO(8)$&57&10.5\%\\
\hline
$SO(10)$&74&13.63\%\\
\hline
$SO(12)$&35&6.446\%\\
\hline
$SO(14)$&72&13.26\%\\
\hline
$E_6$&39&7.182\%\\
\hline
$E_7$&10&1.842\%\\
\hline \hline
\end{tabular}
\label{tab: SU5_NAHE_O3L1_Gauge_Groups}
\end{center}
\end{table}
The most striking feature of this data set is that there are models with three chiral generations both with and without hidden sector duplicates.
The significance of this finding is that these models do not have rank cuts, and thus carry a geometric interpretation.
This implies that these models may be written in another construction method, particularly that of orbifolding \cite{Donagi:2008}.
These are the first three-generation models of their kind.
More analysis will be done in Sec.\ \ref{sec: Three_Generation_Models_With_a_Geometric_Interpretation} regarding this new class of three-generation models.
Other statistics for this data set are plotted in Figure \ref{fig: SU5_NAHE_O3L1_Statistics}.
\subsection{Pati-Salam Models}
Pati-Salam models consist of models with a gauge group $SO(6)\otimes SO(4)$, which is isomorphic to the gauge group $SU(4)\otimes SU(2)\otimes SU(2)$.
The latter is the form of this gauge group which appears in WCFFHS models.
The quark and lepton generations are in representations of (4,2,1), while the anti-quarks and anti-leptons are in representations of ($\overline{4}$,1,2).
Because these data sets are formed by examining all permutations of possible observable sectors, the same statistics will emerge when the chiral generations and anti-generations are examined separately. 
As there are no three-generation models in this data set, this case will not be considered.
The equation for the number of net chiral generations is 
\begin{equation}
|N_{(4,2,1)}-N_{(\overline{4},2,1)}|.\label{eqn: PS_Net_Generations}
\end{equation}
There are 243 unique models containing this gauge group in the order-2 NAHE extensions, and there are 1,648 unique models with this gauge group in the order-3 NAHE extensions.
The abundance of these models is expected; as they contain the most common non-Abelian gauge group, $SU(2)$, along with a group that the NAHE set model already has.

No order-2 models with the Pati-Salam gauge group have net chiral matter generations, but other statistics related to those models are presented, as are
order-3 model statistics.
For the order-2 models, the hidden sector gauge group content is given in Table \ref{tab: PS_NAHE_O2L1_Gauge_Groups}. 
In this data set, both $SU(2)$ at KM-level 1 and $SU(2)^{(2)}$ at KM-level 2 are included as possibilities for the observable Pati-Salam gauge group.
The number of charged exotics is presented in Figure \ref{fig: PS_NAHE_O2L1_Exotics}.
The hidden sector gauge group content for the order-3 models is presented in Table \ref{tab: PS_NAHE_O3L1_Gauge_Groups}.
The number of chiral generations and observable sector charged exotics for these models are plotted in Figure \ref{fig: PS_NAHE_O3L1_OS_Statistics}.

\begin{table}
\caption{The hidden sector gauge group content in Pati-Salam models from the NAHE + O2L1 data set.}
\begin{center}
\begin{tabular}{||c|c|c||}
\hline \hline
Gauge Group & Number of Unique Models & \% of Unique Models \\
\hline \hline
$SO(5)$&60&24.69\%\\
\hline
$SO(8)$&51&20.99\%\\
\hline
$SO(10)$&61&25.1\%\\
\hline
$SO(12)$&1&0.4115\%\\
\hline
$SO(16)$&80&32.92\%\\
\hline
$SO(20)$&2&0.823\%\\
\hline
$E_7$&86&35.39\%\\
\hline
$E_8$&75&30.86\%\\
\hline
$U(1)$&185&76.13\%\\
\hline \hline
\end{tabular}
\label{tab: PS_NAHE_O2L1_Gauge_Groups}
\end{center}
\end{table}
\begin{figure}
\begin{center}
\begin{tikzpicture}
\begin{axis} [ybar, ylabel = Number of Distinct Models,  xlabel = Number of Charged Exotics]
\addplot[draw=black, fill=black]coordinates{
(4,6)(6,8)(7,4)(8,52)(9,8)(10,42)(11,22)(12,140)(13,72)(14,144)(15,190)(16,174)(17,154)(18,228)(19,230)(20,212)(21,292)(22,248)(23,346)(24,254)(25,204)(26,258)(27,312)(28,210)(29,214)(30,120)(31,152)(32,144)(33,186)(34,140)(35,80)(36,48)(37,100)(38,60)(39,56)(40,68)(41,16)(42,42)(43,6)(44,62)(45,48)(46,24)(47,88)(48,28)(49,40)(50,36)(51,40)(52,22)(53,8)(54,52)(55,40)(56,24)(58,6)(59,62)(60,32)(61,16)(66,24)(70,24)};
\end{axis}
\end{tikzpicture}
\caption{The number of observable sector charged exotics from Pati-Salam models in the NAHE + O2L1 data set.}
\label{fig: PS_NAHE_O2L1_Exotics}
\end{center}
\end{figure}
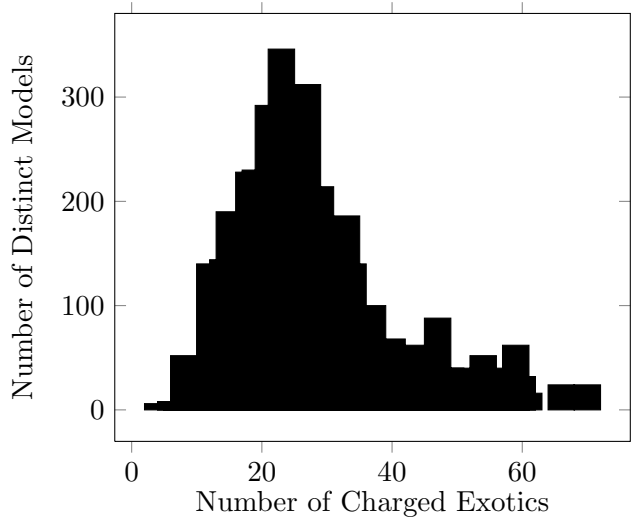
\begin{table}
\caption{The hidden sector gauge group content of the Pati-Salam models in the NAHE + O3L1 data set.}
\begin{center}
\begin{tabular}{||c|c|c||}
\hline \hline
Gauge Group & Number of Unique Models & \% of Unique Models \\
\hline \hline
$SU(3)$&344&20.87\%\\
\hline
$SU(5)$&174&10.56\%\\
\hline
$SU(6)$&414&25.12\%\\
\hline
$SU(7)$&116&7.039\%\\
\hline
$SU(8)$&214&12.99\%\\
\hline
$SU(9)$&30&1.82\%\\
\hline
$SU(10)$&31&1.881\%\\
\hline
$SU(11)$&5&0.3034\%\\
\hline
$SU(12)$&24&1.456\%\\
\hline
$SO(8)$&422&25.61\%\\
\hline
$SO(10)$&332&20.15\%\\
\hline
$SO(12)$&255&15.47\%\\
\hline
$SO(14)$&163&9.891\%\\
\hline
$SO(16)$&106&6.432\%\\
\hline
$SO(20)$&33&2.002\%\\
\hline
$E_6$&81&4.915\%\\
\hline
$E_7$&56&3.398\%\\
\hline
$E_8$&15&0.9102\%\\
\hline
$U(1)$&1615&98\%\\
\hline \hline
\end{tabular}
\label{tab: PS_NAHE_O3L1_Gauge_Groups}
\end{center}
\end{table}
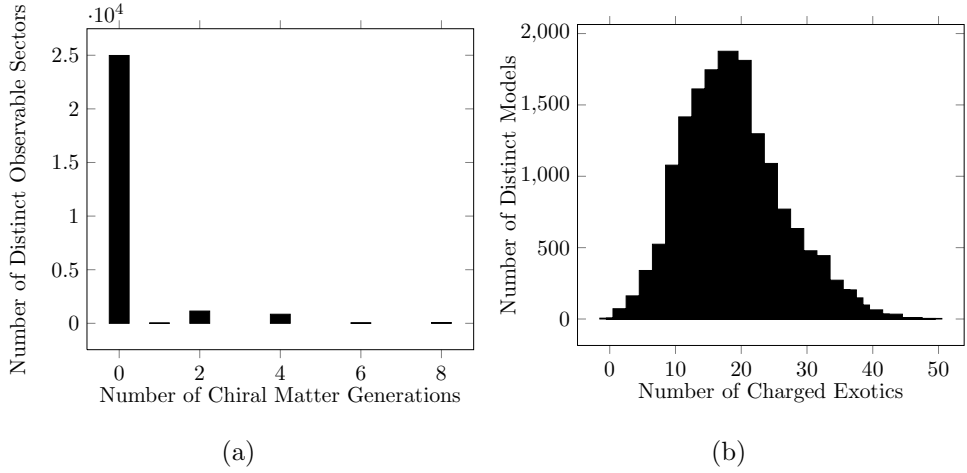
\begin{figure}
\begin{center}
\subfloat[][]{
\begin{tikzpicture}[scale=0.75]
\begin{axis} [ybar, ylabel = Number of Distinct Observable Sectors,  xlabel = Number of Chiral Matter Generations]
\addplot[draw=black, fill=black]coordinates{
(0,24980)(1,36)(2,1144)(4,849)(6,48)(8,67)};
\end{axis}
\end{tikzpicture}
}
\subfloat[][]{
\begin{tikzpicture}[scale=0.75]
\begin{axis} [ybar, ylabel = Number of Distinct Models,  xlabel = Number of Charged Exotics]
\addplot[draw=black, fill=black]coordinates{
(0,6)(1,6)(2,72)(3,56)(4,162)(5,80)(6,340)(7,232)(8,524)(9,438)(10,1078)(11,764)(12,1416)(13,974)(14,1612)(15,1112)(16,1746)(17,1088)(18,1876)(19,1426)(20,1812)(21,1260)(22,1298)(23,896)(24,1090)(25,756)(26,770)(27,626)(28,634)(29,458)(30,478)(31,258)(32,444)(33,164)(34,272)(35,204)(36,204)(37,148)(38,98)(39,56)(40,64)(41,34)(42,18)(43,34)(44,10)(45,10)(46,10)(47,2)(48,4)(49,4)};
\end{axis}
\end{tikzpicture}
}
\caption{Statistics related to observable matter in the Pati-Salam models from the NAHE + O3L1 data set.}
\label{fig: PS_NAHE_O3L1_OS_Statistics}
\end{center}
\end{figure}
Note that the number of chiral generations is zero in most cases for the NAHE + O3L1 data set as well. 
This implies that the symmetry breaking to the two $SU(2)$ gauge groups splits the three distinct generations of the NAHE observable sector evenly.
Thus, three generation Pati-Salam models likely require more complicated basis vector sets.
The remaning statistics are presented in Figure \ref{fig: PS_NAHE_O2L1_Statistics}.
Statistics for the NAHE + O3L1 data set are presented in Figure \ref{fig: PS_NAHE_O3L1_Statistics}.
\begin{figure}
\begin{center}
\subfloat[][]{
\begin{tikzpicture}[scale=0.75]
\begin{axis} [ybar, ylabel = Number of Distinct Models, xlabel = Number of Gauge Group Factors]
\addplot[draw=black, fill=black]coordinates{
(6,4) (7,48) (8,45) (9,64) (10,37) (11,24) (12,9) (13,7) (14,5) };
\end{axis}
\end{tikzpicture}
}
\subfloat[][]{
\begin{tikzpicture}[scale=0.75]
\begin{axis} [ybar, ylabel = Number of Distinct Models, xlabel = Number of $U(1)$ Factors]
\addplot[draw=black, fill=black]coordinates{
(0,58) (1,85) (2,69) (3,31) };
\end{axis}
\end{tikzpicture}
}\\
\subfloat[][]{
\begin{tikzpicture}[scale=0.75]
\begin{axis} [nodes near coords, ybar, ylabel = Number of Distinct Models, xlabel = Number of ST SUSYs]
\addplot[draw=black, fill=black]coordinates{
(0,129) (1,114) (2, 0) (4, 0) };
\end{axis}
\end{tikzpicture}
}
\subfloat[][]{
\begin{tikzpicture}[scale=0.75]
\begin{axis} [ybar, ylabel = Number of Distinct Models,  xlabel = Number of NA Singlets]
\addplot[draw=black, fill=black]coordinates{
(0,107)(2,21)(4,63)(6,19)(8,29)(12,4)};
\end{axis}
\end{tikzpicture}
}
\caption{Statistics for the Pati-Salam models in the NAHE + O2L1 data set.}
\label{fig: PS_NAHE_O2L1_Statistics}
\end{center}
\end{figure}
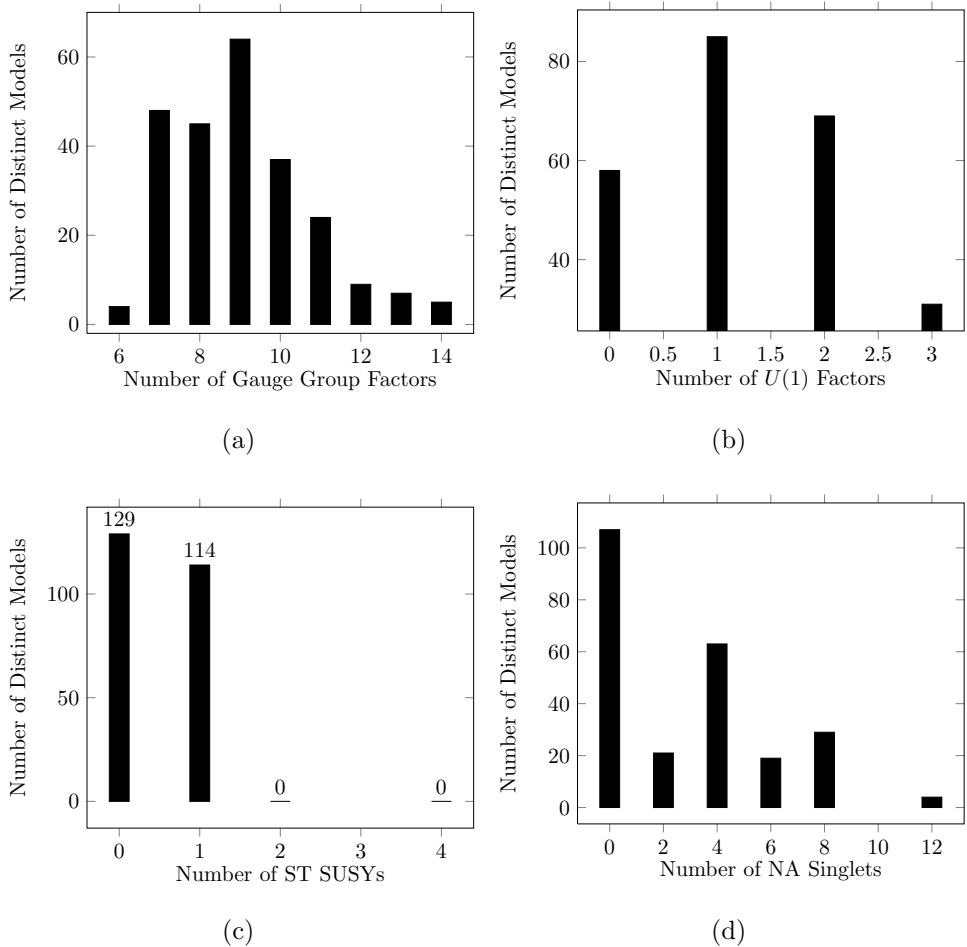
\begin{figure}
\begin{center}
\subfloat[][]{
\begin{tikzpicture}[scale=0.75]
\begin{axis} [ybar, ylabel = Number of Distinct Models, xlabel = Number of Gauge Group Factors]
\addplot[draw=black, fill=black]coordinates{
(6,7) (7,98) (8,144) (9,149) (10,340) (11,267) (12,327) (13,150) (14,134) (15,20) (16,12) };
\end{axis}
\end{tikzpicture}
}
\subfloat[][]{
\begin{tikzpicture}[scale=0.75]
\begin{axis} [ybar, ylabel = Number of Distinct Models, xlabel = Number of $U(1)$ Factors]
\addplot[draw=black, fill=black]coordinates{
(0,33) (1,157) (2,333) (3,178) (4,436) (5,261) (6,94) (7,128) (8,20) (9,8) };
\end{axis}
\end{tikzpicture}
}\\
\subfloat[][]{
\begin{tikzpicture}[scale=0.75]
\begin{axis} [nodes near coords, ybar, ylabel = Number of Distinct Models, xlabel = Number of ST SUSYs]
\addplot[draw=black, fill=black]coordinates{
(0,794) (1,730) (2,124) (4, 0) };
\end{axis}
\end{tikzpicture}
}
\subfloat[][]{
\begin{tikzpicture}[scale=0.75]
\begin{axis} [ybar, ylabel = Number of Distinct Models,  xlabel = Number of NA Singlets]
\addplot[draw=black, fill=black]coordinates{
(0,711)(1,10)(2,158)(3,12)(4,322)(6,89)(8,150)(10,29)(12,62)(14,11)(16,30)(18,8)(20,14)(22,8)(24,13)(26,4)(28,1)(30,4)(32,5)(36,5)(38,2)};
\end{axis}
\end{tikzpicture}
}
\caption{Statistics for the Pati-Salam models in the NAHE + O3L1 data set.}
\label{fig: PS_NAHE_O3L1_Statistics}
\end{center}
\end{figure}
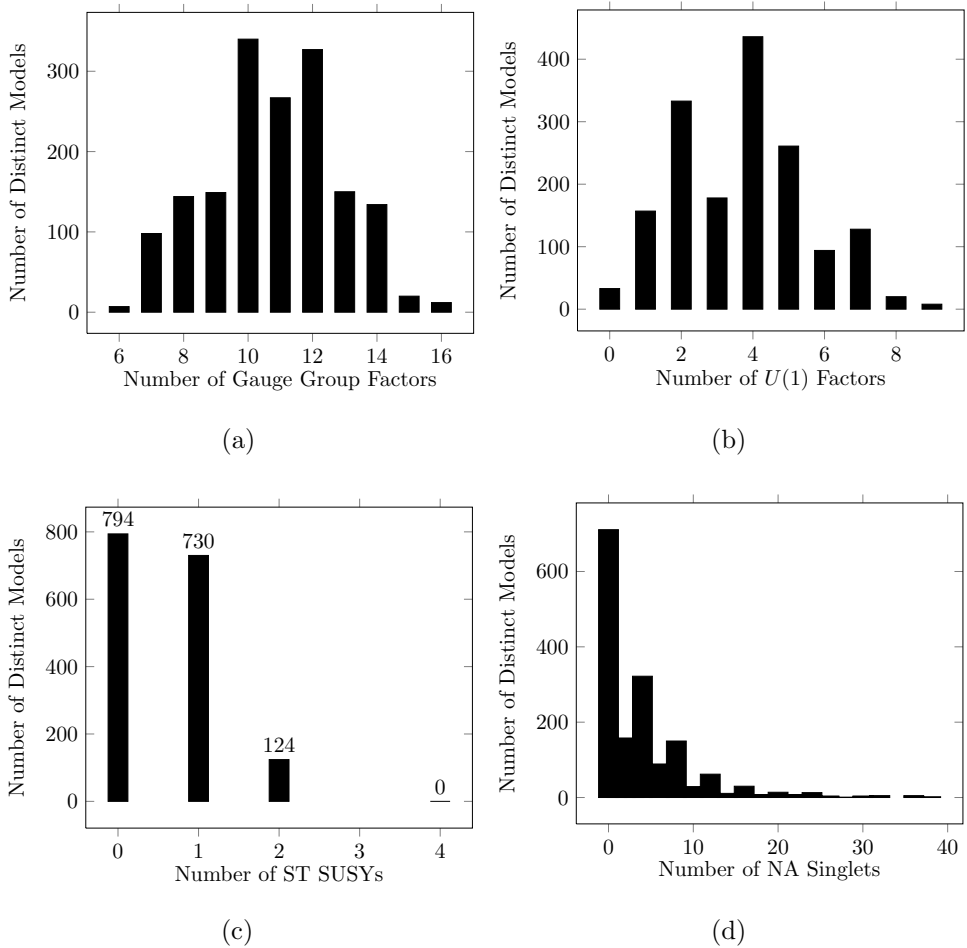
\subsection{Left-Right Symmetric Models}
The final GUT considered in this study is a derivative of the Pati-Salam GUT group referred to as the Left-Right Symmetric group.
It retails the dual-$SU(2)$ nature of the Pati-Salam GUT, but the $SU(4)$ gauge group is broken into an $SU(3)$ group directly representing the strong force.
The generations of quarks fit into a (3,2,1)-dimensional representation while the generations of anti-quarks fit into a ($\overline{3}$,1,2)-dimensional representation.
The lepton and anti-lepton generations are placed in a (1,2,1) and (1,1,2) representation, respectively.
As the quark generations are usually more constraining in WCFFHS models, the term chiral matter generation refers only to the quarks, while the term chiral anti-generation refers only to the anti-quarks.
Lepton generations will need to be taken into account when considering a quasi-realistic model, but here statistics will be gathered only with respect to the quark generations for simplicity.

\begin{figure}
\begin{center}
\subfloat[][]{
\begin{tikzpicture}[scale=0.75]
\begin{axis} [ybar, ylabel = Number of Distinct Observable Sectors,  xlabel = Number of Chiral Matter Generations]
\addplot[draw=black, fill=black]coordinates{
(0,7392)(1,1246)(2,788)(3,70)(4,284)(5,20)(6,32)(8,6)};
\end{axis}
\end{tikzpicture}
}
\subfloat[][]{
\begin{tikzpicture}[scale=0.75]
\begin{axis} [ybar, ylabel = Number of Distinct Models,  xlabel = Number of Charged Exotics]
\addplot[draw=black, fill=black]coordinates{
(0,2)(2,2)(4,48)(5,30)(6,48)(7,56)(8,86)(9,102)(10,176)(11,130)(12,196)(13,170)(14,330)(15,186)(16,456)(17,366)(18,544)(19,526)(20,500)(21,534)(22,610)(23,552)(24,614)(25,564)(26,448)(27,408)(28,384)(29,330)(30,266)(31,170)(32,130)(33,172)(34,184)(35,86)(36,136)(37,68)(38,74)(39,26)(40,38)(41,24)(42,32)(43,4)(44,10)(45,6)(46,8)(49,2)(53,4)};
\end{axis}
\end{tikzpicture}
}
\caption{Observable matter statistics for the Left-Right Symmetric models in the NAHE + O3L1 data set.}
\label{fig: LRSym_NAHE_O3L1_OS_Statistics}
\end{center}
\end{figure}
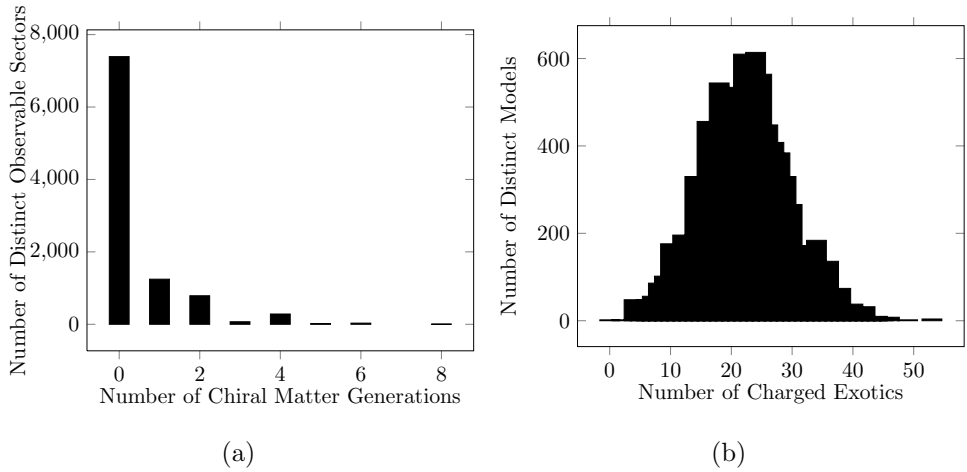

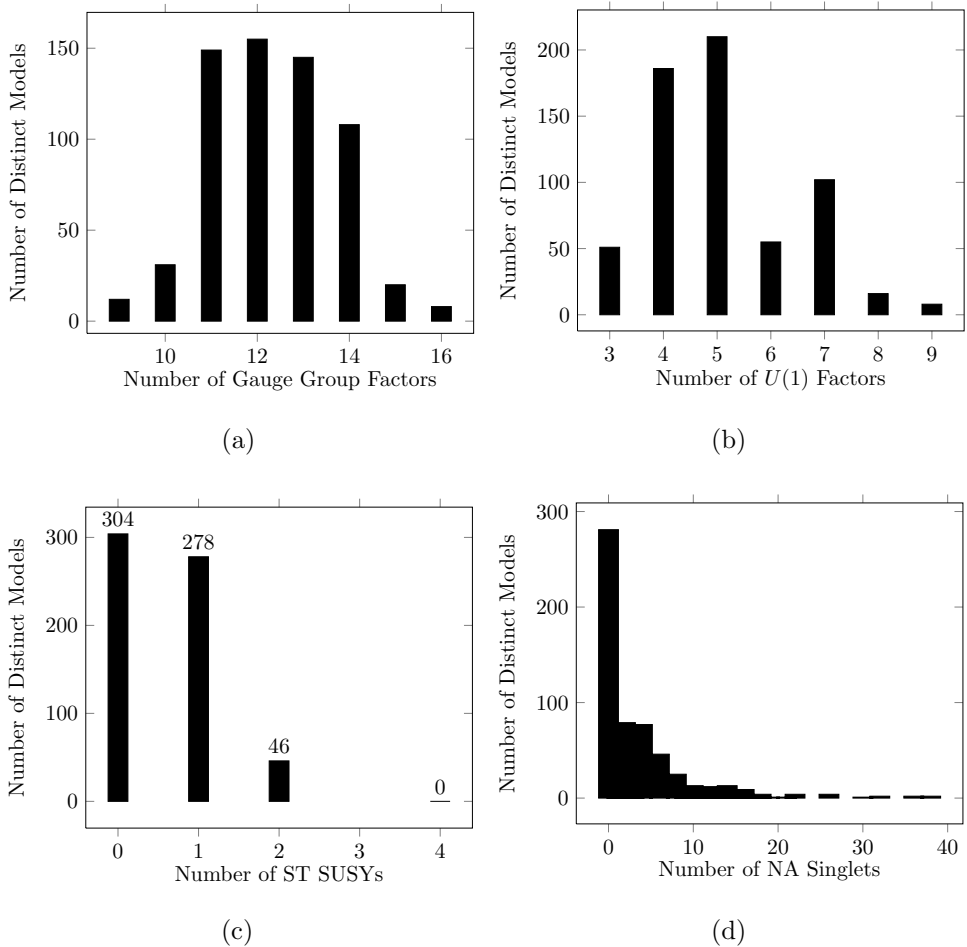
\begin{figure}
\begin{center}
\subfloat[][]{
\begin{tikzpicture}[scale=0.75]
\begin{axis} [ybar, ylabel = Number of Distinct Models, xlabel = Number of Gauge Group Factors]
\addplot[draw=black, fill=black]coordinates{
(9,12) (10,31) (11,149) (12,155) (13,145) (14,108) (15,20) (16,8) };
\end{axis}
\end{tikzpicture}
}
\subfloat[][]{
\begin{tikzpicture}[scale=0.75]
\begin{axis} [ybar, ylabel = Number of Distinct Models, xlabel = Number of $U(1)$ Factors]
\addplot[draw=black, fill=black]coordinates{
(3,51) (4,186) (5,210) (6,55) (7,102) (8,16) (9,8) };
\end{axis}
\end{tikzpicture}
}\\
\subfloat[][]{
\begin{tikzpicture}[scale=0.75]
\begin{axis} [nodes near coords, ybar, ylabel = Number of Distinct Models, xlabel = Number of ST SUSYs]
\addplot[draw=black, fill=black]coordinates{
(0,304) (1,278) (2,46) (4, 0) };
\end{axis}
\end{tikzpicture}
}
\subfloat[][]{
\begin{tikzpicture}[scale=0.75]
\begin{axis} [ybar, ylabel = Number of Distinct Models,  xlabel = Number of NA Singlets]
\addplot[draw=black, fill=black]coordinates{
(0,281)(1,10)(2,79)(3,32)(4,77)(6,46)(8,25)(9,2)(10,13)(11,1)(12,12)(13,3)(14,13)(15,2)(16,9)(17,2)(18,4)(19,1)(21,1)(22,4)(26,4)(30,1)(32,2)(36,2)(38,2)};
\end{axis}
\end{tikzpicture}
}
\caption{Statistics for the Left-Right Symmetric models in the NAHE + O3L1 data set.}
\label{fig: LRSym_NAHE_O3L1_Statistics}
\end{center}
\end{figure}

The number of net chiral (anti)generations is given by 
\begin{equation}
|N_{(3,2,1)} - N_{(\overline{3},2,1)}|.\label{eqn: LRSym_Net_Generations}
\end{equation}
Since the statistics loop over all possible observable sector configurations, the statistical data on the net number of chiral generations and anti-generations are identical.
There are no models with the gauge group in the NAHE + O2L1 data set, as that data set contains no models with $SU(3)$ gauge groups.
There are 628 distinct models in the NAHE + O3L1 data set with this gauge group.
The hidden sector gauge content of those models is presented in Table \ref{tab: LRSym_NAHE_O3L1_Gauge_Groups}. 
The number of net chiral generations is presented in Figure \ref{fig: LRSym_NAHE_O3L1_OS_Statistics} along with the number of observable sector charged exotics.
\begin{table}
\caption{The hidden sector gauge group content of the Left-Right Symmetric models in the NAHE + O3L1 data set.}
\begin{center}
\begin{tabular}{||c|c|c||}
\hline \hline
Gauge Group & Number of Unique Models & \% of Unique Models \\
\hline \hline
$SU(4)$&344&54.78\%\\
\hline
$SU(5)$&264&42.04\%\\
\hline
$SU(6)$&112&17.83\%\\
\hline
$SU(7)$&149&23.73\%\\
\hline
$SU(8)$&84&13.38\%\\
\hline
$SU(9)$&29&4.618\%\\
\hline
$SU(10)$&10&1.592\%\\
\hline
$SU(11)$&17&2.707\%\\
\hline
$SO(8)$&89&14.17\%\\
\hline
$SO(10)$&95&15.13\%\\
\hline
$SO(14)$&53&8.439\%\\
\hline
$E_6$&41&6.529\%\\
\hline
$U(1)$&628&100\%\\
\hline \hline
\end{tabular}
\label{tab: LRSym_NAHE_O3L1_Gauge_Groups}
\end{center}
\end{table}
Like the $SU(5)\otimes U(1)$ data set, there are three-generation models present here. 
There are 70 models with three net chiral generations.
One such model will be presented as an example at the end of the paper.
The remaining statistical information on these models is presented in Figure \ref{fig: LRSym_NAHE_O3L1_Statistics}.
\subsection{MSSM-like Models}
The MSSM\footnote{Here MSSM refers only to the gauge group content. Models with this gauge group may or may not have ST SUSY.} gauge group is $SU(3)\otimes SU(2)\otimes U(1)$.
A generation of quarks fit in a (3,2) representation of these groups.
The leptons fit in a (1,2) representation.
The generations of antimatter are charged differently, as the antiparticles do not have isospin.
A generation of antimatter consists of two ($\overline{3}$,1) representations, one for the ``up"-type quarks and one for the ``down"-type quarks.
While the leptons fit into a (1,2) representation, the anti-leptons are (1,1) singlets.
As was the case with Left-Right Symmetric models, the terms chiral generation and anti-generation refer only to the quarks.
While the lepton generations must also be considered, statistics are only gathered for the quark generations, as they are more constraining.

\begin{figure}
\begin{center}
\subfloat[][Quarks]{
\begin{tikzpicture}[scale=0.75]
\begin{axis} [ybar, ylabel = Number of Distinct Observable Sectors,  xlabel = Number of Chiral Matter Generations]
\addplot[draw=black, fill=black]coordinates{
(0,2876)(1,641)(2,372)(3,34)(4,82)(5,8)(6,8)(8,2)};
\end{axis}
\end{tikzpicture}
}
\subfloat[][Anti-quarks]{
\begin{tikzpicture}[scale=0.75]
\begin{axis} [ybar, ylabel = Number of Distinct Observable Sectors,  xlabel = Number of Chiral Matter Generations]
\addplot[draw=black, fill=black]coordinates{
(0,3404)(1,131)(2,214)(3,87)(4,99)(5,24)(6,32)(7,16)(8,4)(10,6)(12,6)};
\end{axis}
\end{tikzpicture}
}\\
\subfloat[][]{
\begin{tikzpicture}[scale=0.75]
\begin{axis} [ybar, ylabel = Number of Distinct Models,  xlabel = Number of Charged Exotics]
\addplot[draw=black, fill=black]coordinates{
(0,4)(2,2)(3,1)(4,33)(5,9)(6,50)(7,91)(8,132)(9,104)(10,179)(11,89)(12,270)(13,147)(14,220)(15,184)(16,292)(17,228)(18,275)(19,172)(20,281)(21,156)(22,225)(23,160)(24,187)(25,89)(26,110)(27,40)(28,86)(29,46)(30,59)(31,12)(32,36)(33,10)(34,20)(35,2)(36,16)(37,4)(44,1)(46,1)};
\end{axis}
\end{tikzpicture}
}
\caption{Observable matter related statistics for the MSSM models in the NAHE + O3L1 data set.}
\label{fig: MSSM_NAHE_O3L1_OS_Statistics}
\end{center}
\end{figure}
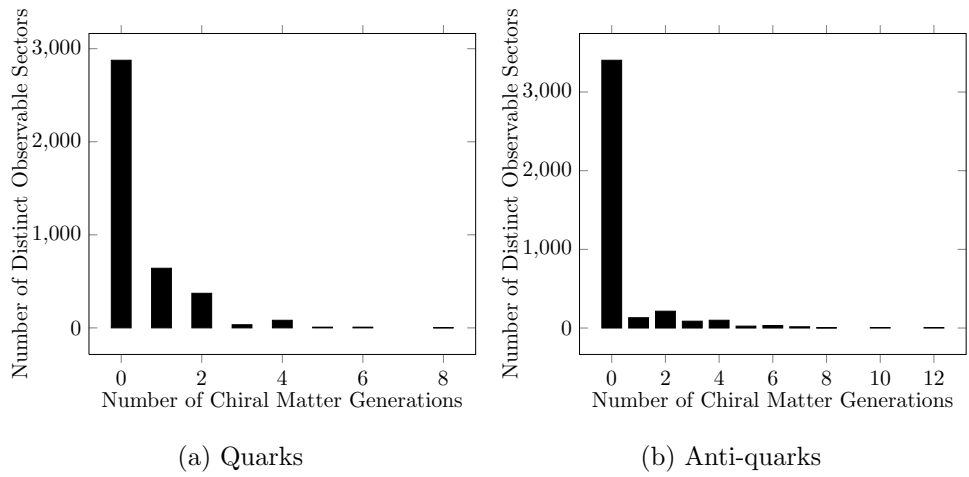
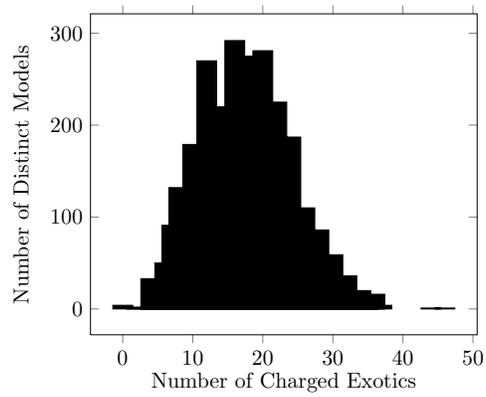

\begin{figure}
\begin{center}
\subfloat[][]{
\begin{tikzpicture}[scale=0.75]
\begin{axis} [ybar, ylabel = Number of Distinct Models, xlabel = Number of Gauge Group Factors]
\addplot[draw=black, fill=black]coordinates{
(9,19) (10,67) (11,228) (12,172) (13,153) (14,108) (15,20) (16,8) };
\end{axis}
\end{tikzpicture}
}
\subfloat[][]{
\begin{tikzpicture}[scale=0.75]
\begin{axis} [ybar, ylabel = Number of Distinct Models, xlabel = Number of $U(1)$ Factors]
\addplot[draw=black, fill=black]coordinates{
(3,51) (4,216) (5,282) (6,92) (7,110) (8,16) (9,8) };
\end{axis}
\end{tikzpicture}
}\\
\subfloat[][]{
\begin{tikzpicture}[scale=0.75]
\begin{axis} [nodes near coords, ybar, ylabel = Number of Distinct Models, xlabel = Number of ST SUSYs]
\addplot[draw=black, fill=black]coordinates{
(0,376) (1,345) (2,54) (4, 0) };
\end{axis}
\end{tikzpicture}
}
\subfloat[][]{
\begin{tikzpicture}[scale=0.75]
\begin{axis} [ybar, ylabel = Number of Distinct Models,  xlabel = Number of NA Singlets]
\addplot[draw=black, fill=black]coordinates{
(0,313)(1,10)(2,94)(3,34)(4,112)(5,2)(6,57)(7,12)(8,51)(9,2)(10,15)(11,1)(12,14)(13,7)(14,15)(15,2)(16,11)(17,2)(18,4)(19,1)(21,1)(22,4)(26,4)(30,1)(32,2)(36,2)(38,2)};
\end{axis}
\end{tikzpicture}
}
\caption{Statistics for MSSM models in the NAHE + O3L1 data set.}
\label{fig: MSSM_NAHE_O3L1_Statistics}
\end{center}
\end{figure}
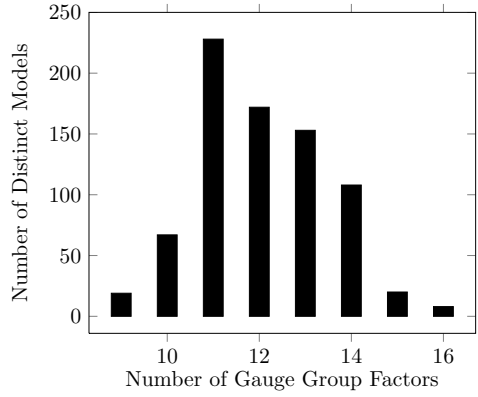
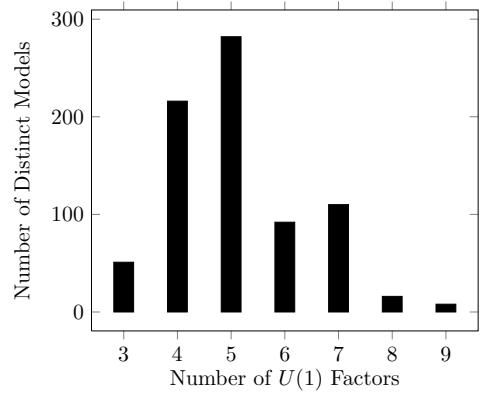
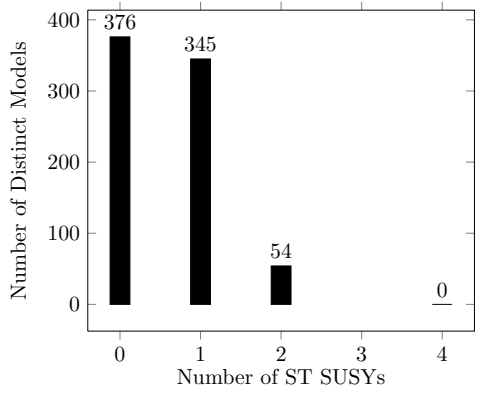
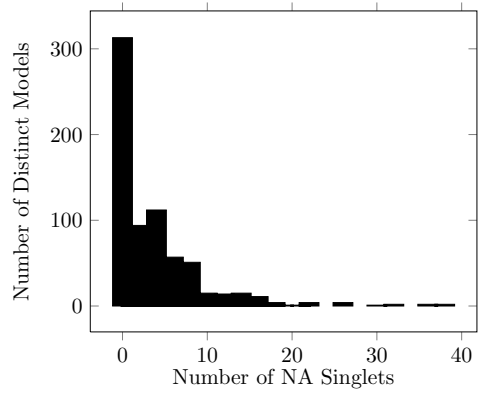

The equation for the number of net chiral matter generations is 
\begin{equation}
|N_{(3,2)}-N_{(\overline{3},2)}|,\label{eqn: MSSM_Net_Generations}
\end{equation}
while the number of net chiral antimatter generations is 
\begin{equation}
|N_{(3,1)}-N_{(\overline{3},1)}|.\label{eqn: MSSM_Net_Anti_Generations}
\end{equation}
There are no models with this gauge group from the NAHE + O2L1 data set since there are no $SU(3)$ gauge groups.
There are, however, 775 models in the NAHE + O3L1 data set with the MSSM group.
The hidden sector gauge group content of models containing the MSSM gauge group is presented in Table \ref{tab: MSSM_NAHE_O3L1_Gauge_Groups}.
The number of net chiral generations and anti-generations are presented in Figure \ref{fig: MSSM_NAHE_O3L1_OS_Statistics}.
The number of observable sector charged exotics is also plotted in Figure \ref{fig: MSSM_NAHE_O3L1_OS_Statistics}.
\begin{table}
\caption{The hidden sector gauge group content of the MSSM models in the NAHE + O3L1 data set.}
\begin{center}
\begin{tabular}{||c|c|c||}
\hline \hline
Gauge Group & Number of Unique Models & \% of Unique Models \\
\hline \hline
$SU(4)$&412&53.16\%\\
\hline
$SU(5)$&374&48.26\%\\
\hline
$SU(6)$&112&14.45\%\\
\hline
$SU(7)$&169&21.81\%\\
\hline
$SU(8)$&112&14.45\%\\
\hline
$SU(9)$&41&5.29\%\\
\hline
$SU(10)$&10&1.29\%\\
\hline
$SU(11)$&17&2.194\%\\
\hline
$SO(8)$&97&12.52\%\\
\hline
$SO(10)$&111&14.32\%\\
\hline
$SO(12)$&35&4.516\%\\
\hline
$SO(14)$&68&8.774\%\\
\hline
$E_6$&46&5.935\%\\
\hline \hline
\end{tabular}
\label{tab: MSSM_NAHE_O3L1_Gauge_Groups}
\end{center}
\end{table}
There are models with three chiral matter generations, as well as models with three anti-generations.
However, none of the models have three generations of quarks and anti-quarks.
While these findings are still significant due to their novelty, they do not point towards phenomenologically realistic models as the $SU(5)\otimes U(1)$ and Left-Right Symmetric models do.
The remaining statistics for these models are presented in Figure \ref{fig: MSSM_NAHE_O3L1_Statistics}.
\subsection{ST SUSYs}
\begin{figure}
\begin{center}
\subfloat[][Full data set] {
\begin{tikzpicture}[scale=0.75]
\begin{axis} [nodes near coords, ybar, ylabel = Number of Distinct Models, xlabel = Number of ST SUSYs]
\addplot[draw=black, fill=black]coordinates{
(0,223) (1,215) (2,1) (4, 0) };
\end{axis}
\end{tikzpicture}
}\\
\subfloat[][$SO(10)$ models]{
\begin{tikzpicture}[scale=0.75]
\begin{axis} [nodes near coords, ybar, ylabel = Number of Distinct Models, xlabel = Number of ST SUSYs]
\addplot[draw=black, fill=black]coordinates{
(0,82) (1,77) (2,1) (4, 0) };
\end{axis}
\end{tikzpicture}
}
\subfloat[][Pati-Salam models]{
\begin{tikzpicture}[scale=0.75]
\begin{axis} [nodes near coords, ybar, ylabel = Number of Distinct Models, xlabel = Number of ST SUSYs]
\addplot[draw=black, fill=black]coordinates{
(0,129) (1,114) (2, 0) (4, 0) };
\end{axis}
\end{tikzpicture}
}
\caption{The distributions of ST SUSYs for the NAHE + O2L1 GUT group data sets.}
\label{fig: NAHE_O2L1_All_ST_SUSYs}
\end{center}
\end{figure}
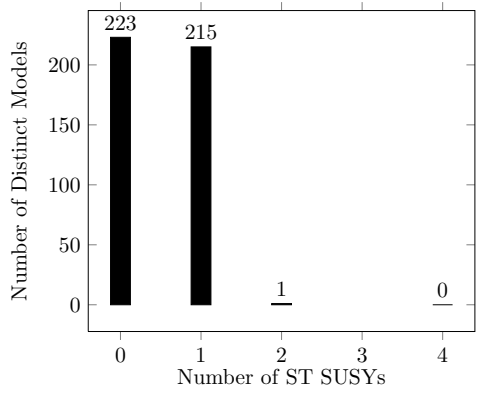
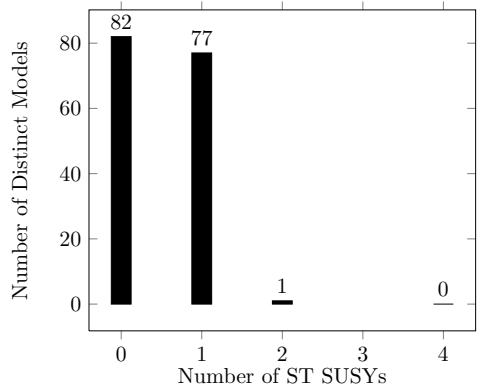
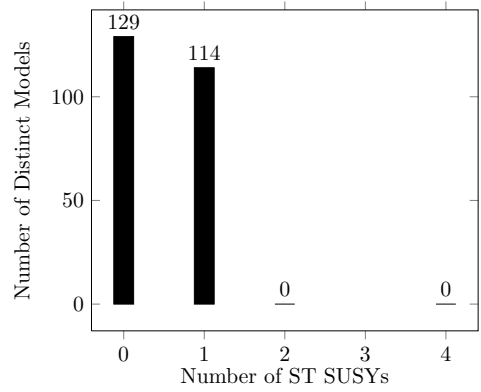

\begin{figure}
\begin{center}
\includegraphics[scale=0.85]{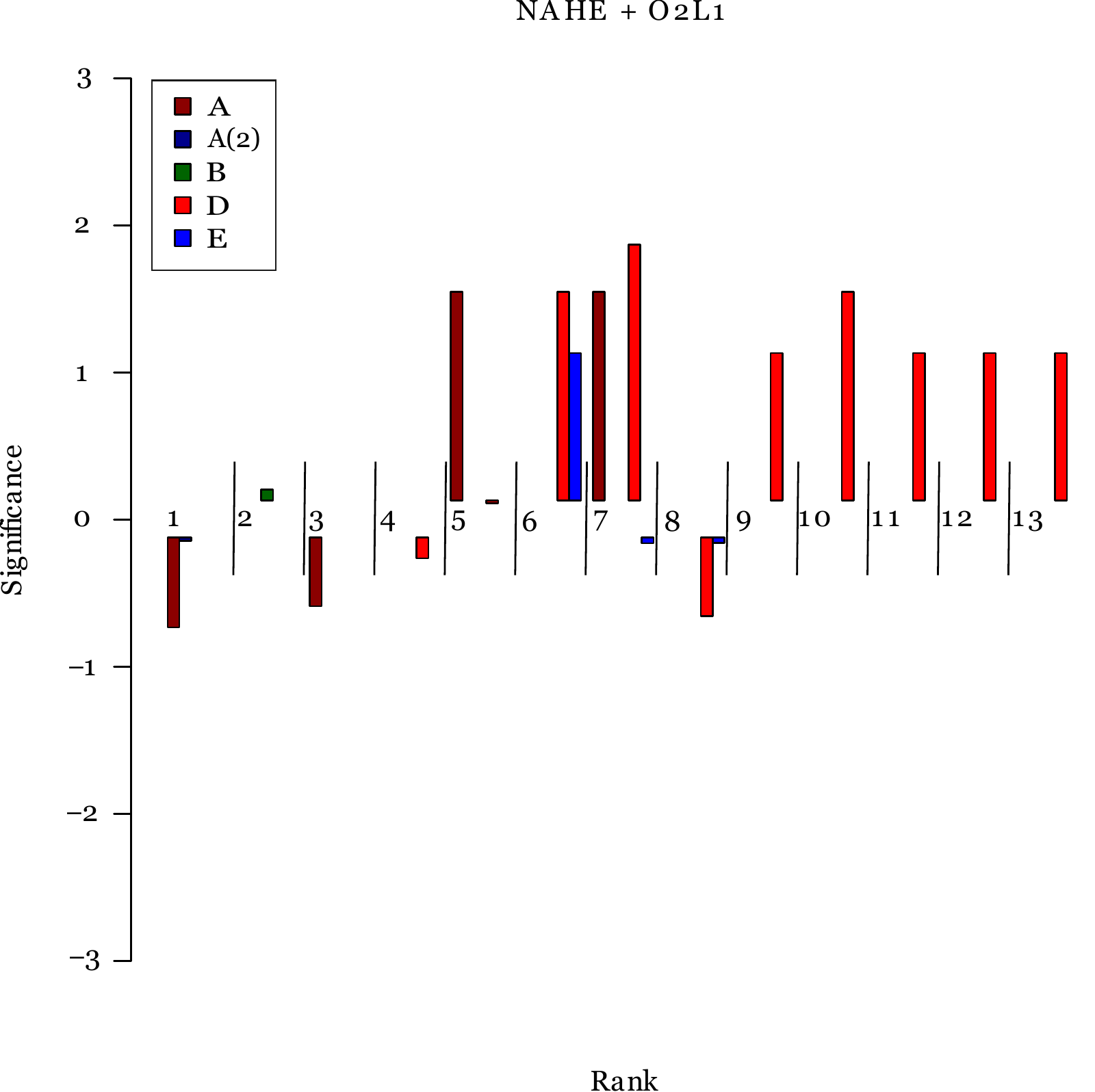}
\caption{The significance values for models in the NAHE + O2L1 data set. Any significance values greater than three indicate a strong statistical significance.}
\label{fig: NAHE_O2L1_ST_SUSY_Significances}
\end{center}
\end{figure}

\begin{figure}
\begin{center}
\includegraphics[scale=0.85]{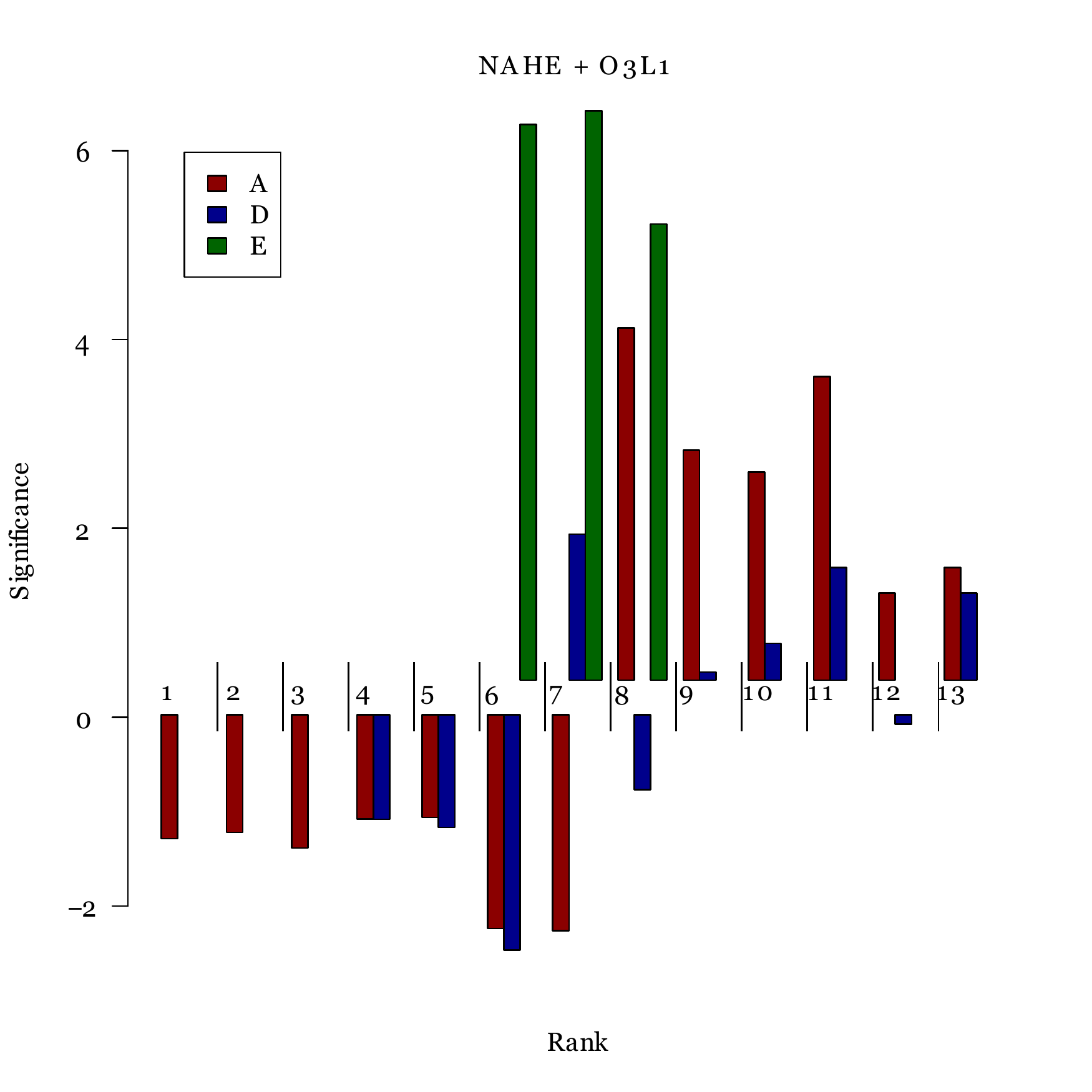}
\caption{The significance values for models in the NAHE + O3L1 data set. Any significance values greater than three indicate a strong statistical significance.}
\label{fig: NAHE_O3L1_ST_SUSY_Significances}
\end{center}
\end{figure}

There is a trend regarding the number of ST SUSYs in GUT models --- that distributions of ST SUSYs for the most part do not change.
Figure \ref{fig: NAHE_O2L1_All_ST_SUSYs} contains the ST SUSY distributions for the full data set, the $SO(10)$ models, and the Pati-Salam models.
It is clear that the ST SUSY distributions are relatively even for each of the sample sets of models. 
The same can be said of order-3 models, whose ST SUSY distributions are presented in Figures \ref{fig: NAHE_O3L1_All_ST_SUSYs} and \ref{fig: NAHE_O3L1_All_ST_SUSYs_B}.
Only the $E_6$ models display any sort of statistical coupling to the number of ST SUSYs, having significantly more N=2 models than any other data set.
The other samples, however, have nearly identical distributions, suggesting that the number of ST SUSYs is not statistically linked to the GUT group content.
\begin{figure}
\begin{center}
\subfloat[][Full data set]{
\begin{tikzpicture}[scale=0.75]
\begin{axis} [nodes near coords, enlargelimits=0.15, ybar, ylabel = Number of Distinct Models, xlabel = Number of ST SUSYs]
\addplot[draw=black, fill=black]coordinates{
(0,1445) (1,1305) (2,286) (4, 0) };
\end{axis}
\end{tikzpicture}
}\\
\subfloat[][$E_6$ models]{
\begin{tikzpicture}[scale=0.75]
\begin{axis} [nodes near coords, ybar, ylabel = Number of Distinct Models, xlabel = Number of ST SUSYs]
\addplot[draw=black, fill=black]coordinates{
(0,81) (1,48) (2,64) (4, 0) };
\end{axis}
\end{tikzpicture}
}
\subfloat[][$SO(10)$ models]{
\begin{tikzpicture}[scale=0.75]
\begin{axis} [nodes near coords, ybar, ylabel = Number of Distinct Models, xlabel = Number of ST SUSYs]
\addplot[draw=black, fill=black]coordinates{
(0,318) (1,294) (2,47) (4, 0) };
\end{axis}
\end{tikzpicture}
}\\
\caption{Some of the distributions of ST SUSYs for the NAHE + O3L1 GUT group data sets.}
\label{fig: NAHE_O3L1_All_ST_SUSYs}
\end{center}
\end{figure}
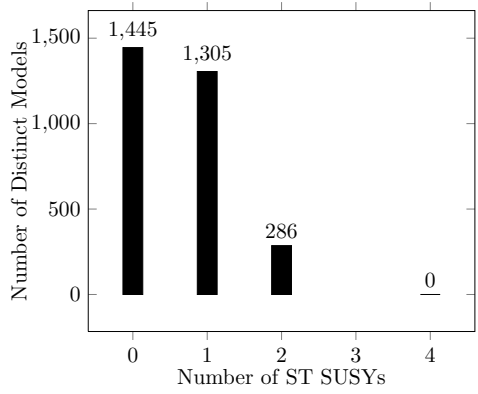
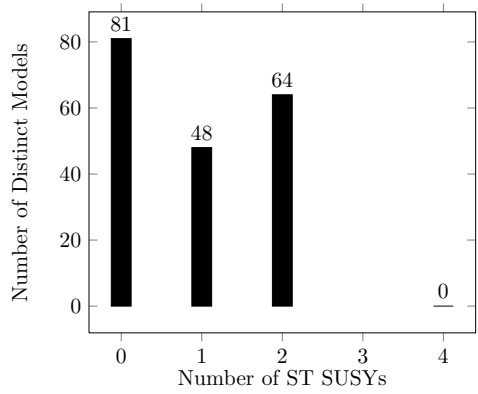
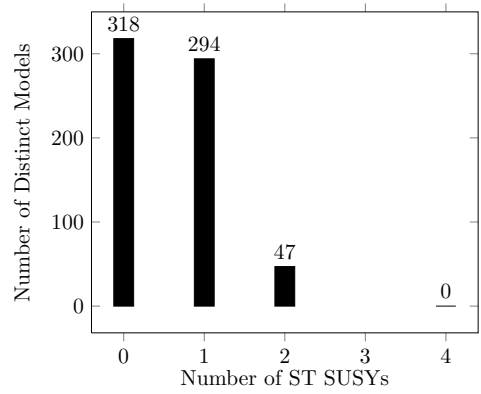
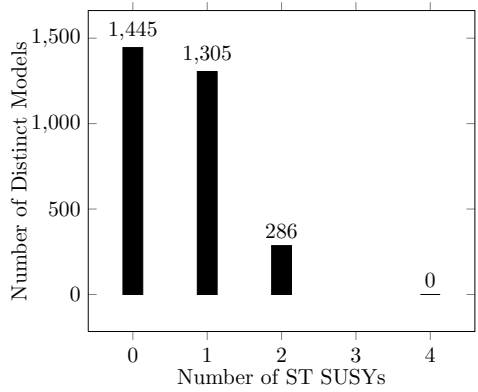
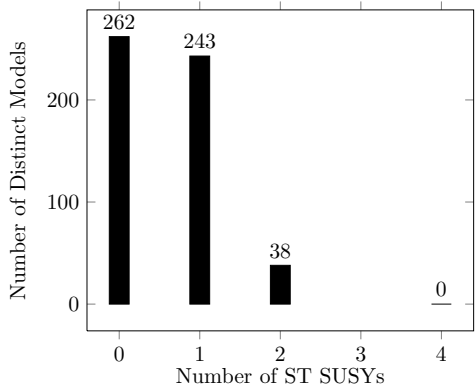
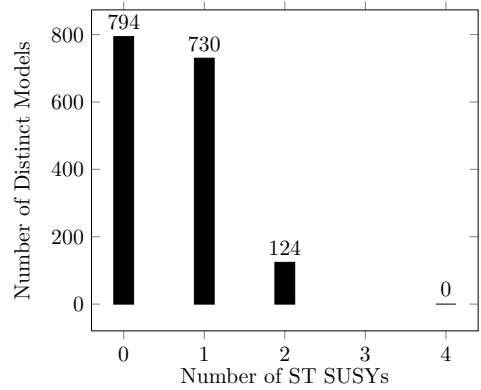
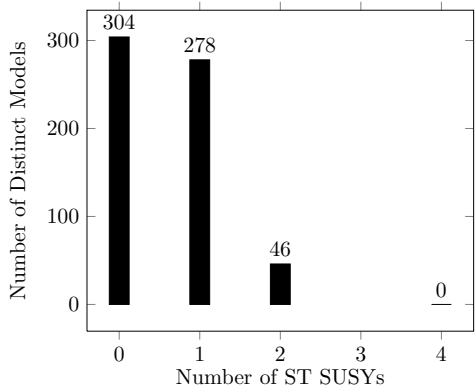
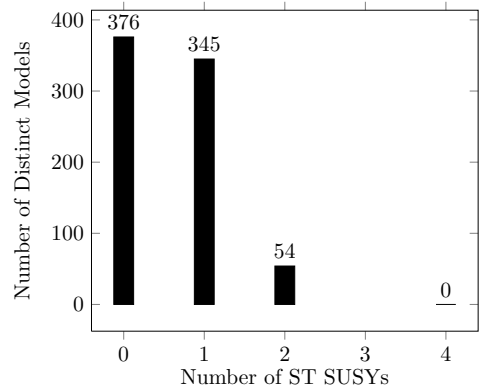
\begin{figure}
\begin{center}
\subfloat[][Full data set]{
\begin{tikzpicture}[scale=0.75]
\begin{axis} [nodes near coords, enlargelimits=0.15, ybar, ylabel = Number of Distinct Models, xlabel = Number of ST SUSYs]
\addplot[draw=black, fill=black]coordinates{
(0,1445) (1,1305) (2,286) (4, 0) };
\end{axis}
\end{tikzpicture}
}\\
\subfloat[][$SU(5)\otimes U(1)$]{
\begin{tikzpicture}[scale=0.75]
\begin{axis} [nodes near coords, ybar, ylabel = Number of Distinct Models, xlabel = Number of ST SUSYs]
\addplot[draw=black, fill=black]coordinates{
(0,262) (1,243) (2,38) (4, 0) };
\end{axis}
\end{tikzpicture}
}
\subfloat[][Pati-Salam models]{
\begin{tikzpicture}[scale=0.75]
\begin{axis} [nodes near coords, ybar, ylabel = Number of Distinct Models, xlabel = Number of ST SUSYs]
\addplot[draw=black, fill=black]coordinates{
(0,794) (1,730) (2,124) (4, 0) };
\end{axis}
\end{tikzpicture}
}\\
\subfloat[][Left-Right Symmetric models]{
\begin{tikzpicture}[scale=0.75]
\begin{axis} [nodes near coords, ybar, ylabel = Number of Distinct Models, xlabel = Number of ST SUSYs]
\addplot[draw=black, fill=black]coordinates{
(0,304) (1,278) (2,46) (4, 0) };
\end{axis}
\end{tikzpicture}
}
\subfloat[][MSSM models]{
\begin{tikzpicture}[scale=0.75]
\begin{axis} [nodes near coords, ybar, ylabel = Number of Distinct Models, xlabel = Number of ST SUSYs]
\addplot[draw=black, fill=black]coordinates{
(0,376) (1,345) (2,54) (4, 0) };
\end{axis}
\end{tikzpicture}
}
\caption{The remaining distributions of ST SUSYs for the NAHE + O3L1 GUT group data sets.}
\label{fig: NAHE_O3L1_All_ST_SUSYs_B}
\end{center}
\end{figure}
Further investigations of these findings show several statistical couplings for higher ST SUSY models containing certain gauge group factors.
The statistical test to be used invokes the Central Limit Theorem, which is applicable to populations which are well behaved, such as the number ST SUSY distributions discussed.
We will average the number of ST SUSYs per model for models containing each of the gauge groups present.
Random sample averages will be close to the average of the population; therefore any sample drawn based on gauge groups which has an average close to the population average indicates that the average number of ST SUSYs is not coupled to the gauge group.
If the average number of ST SUSYs per model for a particular gauge group (for example, $E_6$) is higher than the population (and if the sample is large enough to be significant), then we can conclude that the gauge group content has an effect on the number of ST SUSYs for a significant percentage of models.
These significances are plotted in \autoref{fig: NAHE_O2L1_ST_SUSY_Significances} for the NAHE + O2L1 data set and \autoref{fig: NAHE_O3L1_ST_SUSY_Significances} for the NAHE + O3L1 data set.

While there are no significant gauge groups in the NAHE + O2L1 data set, all three exceptional groups, as well as $SU(9)$ and $SU(12)$, have significant effects on the average number of ST SUSYs.
This is likely due to the most common embeddings of the SUSY sectors causing additional roots in the gauge groups, promoting them from $SU(n+1)$ to $SO(2n)$ or $E_n$.
Additional analysis will be needed to confirm the cause of this significance.
\clearpage
\section{Three Generation Models With a Geometric Interpretation}\label{sec: Three_Generation_Models_With_a_Geometric_Interpretation}
Several models containing three net chiral matter generations with $SU(5)\otimes U(1)$ and Left-Right Symmetric GUT groups were found in the NAHE + O3L1 data set.
As previously mentioned, this finding is novel because these models do not have rank-cuts, and thus have a geometric interpretation.
The usual statistics will be reported for these models, and a potentially realistic model from each gauge group will be presented as an example.
To determine the viability of these models, more phenomenology must be done. 
In particular, finding the $U(1)$ charges and the superpotential would be the first step, then the D- and F-flat directions can be found.
If the flat directions can eliminate the observable sector charged exotic matter simultaneously with the anomalous $U(1)$ charge, then the model could be considered a quasi-realistic model.
\subsection{A Three Generation $SU(5)\otimes U(1)$ Model}
Presented in this section is an explicit example of a NAHE based three-generation $SU(5)\otimes U(1)$ model with $N=1$ ST SUSY.
The gauge group for this model is $SU(3)^2\otimes SU(4)\otimes SU(5)\otimes SU(6)\otimes U(1)^5$.
\begin{table}[h]
\caption{A basis vector and $k_{ij}$ matrix row which produces a three-generation $SU(5)\otimes U(1)$ model.}
\begin{center}
\begin{tabular}{||c|c|c|c|c|c|c|c|c|c|c||}
\hline \hline
Sec&$N_R$&$\psi$&$x^{12}$&$x^{34}$&$x^{56}$&$\overline{\psi}^{~1,...,5}$&$\overline{\eta}^{~1}$&$\overline{\eta}^{~2}$&$\overline{\eta}^{~3}$&$\overline{\phi}^{~1,...,8}$\\
\hline \hline
$\vec{v}$&3&1&1&0&0&0,0,$\frac{2}{3}$,...,$\frac{2}{3}$&$\frac{2}{3}$&0&$\frac{2}{3}$&0,0,$\frac{2}{3},...,\frac{2}{3}$\\
\hline \hline
\end{tabular}
\vspace{5 mm}\\
\begin{tabular}{||c|c|c|c|c||}
\hline \hline
Sec&O&$y^{~1,2}w^{~5,6}||\overline{y}^{~1,2}\overline{w}^{~5,6}$&$y^{~3,...,6}||\overline{y}^{~3,...,6}$&$w^{~1,...,4}||\overline{w}^{~1,...,4}$\\
\hline \hline
$\vec{v}$&3&0,0,1,1$||$0,0,0,0&0,0,0,0$||\frac{2}{3},\frac{2}{3},\frac{2}{3},\frac{2}{3}$&0,0,1,1$||$$\frac{2}{3}$,$\frac{2}{3}$,$\frac{2}{3},\frac{2}{3}$\\
\hline \hline
\end{tabular}
\vspace{3 mm}\\
$k_{\vec{v},j}$ = (0, 1, 1, 0, 1)
\label{tab: 3G_SU5_NAHE_O3L1_Example_BVs}
\end{center}
\end{table}
Table \ref{tab: 3G_SU5_NAHE_O3L1_Example_BVs} shows the basis vectors, and Table \ref{tab: 3G_SU5_NAHE_O3L1_Example_Particles} shows the particle content.
The observable sector matter is tabulated in Table \ref{tab: 3G_SU5_NAHE_O3L1_Example_Observable_Matter}.
\begin{table}
\caption{Particle content for the three-generation $SU(5)\otimes U(1)$ model. This model also has five $U(1)$ groups and $N=1$ ST SUSY.}
\begin{center}
\begin{tabular}{||c|c|c|c|c|c||}
\hline \hline
\textbf{QTY}&$SU(3)$&$SU(3)$&$SU(4)$&$SU(5)$&$SU(7)$\\
\hline \hline
2&$3_{ }$&$1_{ }$&$1_{ }$&$1_{ }$&$1_{ }$\\
\hline
3&$3_{ }$&$1_{ }$&$1_{ }$&$1_{ }$&$7_{ }$\\
\hline
3&$3_{ }$&$3_{ }$&$1_{ }$&$1_{ }$&$1_{ }$\\
\hline
1&$1_{ }$&$\overline{3}$&$1_{ }$&$1_{ }$&$\overline{7}$\\
\hline
1&$1_{ }$&$\overline{3}$&$1_{ }$&$1_{ }$&$1_{ }$\\
\hline
1&$1_{ }$&$\overline{3}$&$1_{ }$&$1_{ }$&$7_{ }$\\
\hline
2&$1_{ }$&$\overline{3}$&$1_{ }$&$5_{ }$&$1_{ }$\\
\hline
6&$1_{ }$&$1_{ }$&$6_{ }$&$1_{ }$&$1_{ }$\\
\hline
2&$1_{ }$&$1_{ }$&$4_{ }$&$1_{ }$&$1_{ }$\\
\hline
1&$1_{ }$&$1_{ }$&$4_{ }$&$5_{ }$&$1_{ }$\\
\hline
2&$1_{ }$&$1_{ }$&$1_{ }$&$\overline{5}$&$1_{ }$\\
\hline
2&$1_{ }$&$1_{ }$&$1_{ }$&$1_{ }$&$\overline{21}$\\
\hline
6&$1_{ }$&$1_{ }$&$1_{ }$&$1_{ }$&$1_{ }$\\
\hline
1&$1_{ }$&$1_{ }$&$1_{ }$&$1_{ }$&$21_{ }$\\
\hline
3&$1_{ }$&$1_{ }$&$1_{ }$&$5_{ }$&$1_{ }$\\
\hline
2&$1_{ }$&$1_{ }$&$\overline{4}$&$1_{ }$&$1_{ }$\\
\hline
1&$1_{ }$&$1_{ }$&$\overline{4}$&$5_{ }$&$1_{ }$\\
\hline
2&$1_{ }$&$3_{ }$&$1_{ }$&$\overline{5}$&$1_{ }$\\
\hline
1&$1_{ }$&$3_{ }$&$1_{ }$&$1_{ }$&$\overline{7}$\\
\hline
2&$1_{ }$&$3_{ }$&$1_{ }$&$1_{ }$&$1_{ }$\\
\hline
1&$\overline{3}$&$\overline{3}$&$1_{ }$&$1_{ }$&$1_{ }$\\
\hline
1&$\overline{3}$&$1_{ }$&$1_{ }$&$\overline{10}$&$1_{ }$\\
\hline
2&$\overline{3}$&$1_{ }$&$1_{ }$&$\overline{5}$&$1_{ }$\\
\hline
1&$\overline{3}$&$1_{ }$&$1_{ }$&$1_{ }$&$\overline{7}$\\
\hline
2&$\overline{3}$&$1_{ }$&$1_{ }$&$1_{ }$&$1_{ }$\\
\hline \hline
\end{tabular}
\label{tab: 3G_SU5_NAHE_O3L1_Example_Particles}
\end{center}
\end{table}
\begin{table}
\caption{Observable sector matter states without hidden sector charges for the three-generation $SU(5)\otimes U(1)$ model.}
\begin{center}
\begin{tabular}{||c|c|c|c|c|c||}
\hline \hline
\textbf{QTY}&$SU(3)$&$SU(3)$&$SU(4)$&$SU(5)$&$SU(7)$\\
\hline \hline
2&$1$&$\overline{3}$&$1$&$5$&$1$\\
\hline
1&1&1&4&5&1\\
\hline
2&$1$&$1$&$1$&$\overline{5}$&$1$\\
\hline
3&$1$&$1$&$1$&$5$&$1$\\
\hline
1&$1$&$1$&$\overline{4}$&$5$&$1$\\
\hline
2&$1$&$3$&$1$&$\overline{5}$&$1$\\
\hline
1&$\overline{3}$&$1$&$1$&$\overline{10}$&$1$\\
\hline
2&$\overline{3}$&$1$&$1$&$\overline{5}$&$1$\\
\hline \hline
\end{tabular}
\label{tab: 3G_SU5_NAHE_O3L1_Example_Observable_Matter}
\end{center}
\end{table}
There are no (10,$\overline{5}$) generations and three ($\overline{10}$,5) generations in this model, giving it three net chiral generations of matter\footnote{Recall that the definition of barred and unbarred representations is arbitrary.}.
However, counting the hidden sector charges as duplicates, there are 14 extra 5's and 8 extra $\overline{5}$'s.
Because of the numerous $U(1)$ charges, this model is ideal for future $U(1)$ and flat direction analysis.

\clearpage
\subsection{A Three Generation Left-Right Symmetric Model}
Presented in this section is an explicitly constructed three-generation Left-Right Symmetric NAHE based model. 
The gauge group for this model is $SU(2)^2\otimes  SU(3)^2\otimes SU(5)\otimes SO(10)\otimes U(1)^7$, and it has N=1 ST SUSY.
The basis vectors for this model are given in Table \ref{tab: 3G_LRSym_NAHE_O3L1_Example_BVs}. 
The particle content of this model is presented in Table \ref{tab: 3G_LRSym_NAHE_O3L1_Example_Particles}, and the observable matter is presented in Table \ref{tab: 3G_LRSym_NAHE_O3L1_Example_Observable_Matter}.

\begin{table}
\caption{A basis vector and $k_{ij}$ matrix row which produces a three-generation Left-Right Symmetric model.}
\begin{center}
\begin{tabular}{||c|c|c|c|c|c|c|c|c|c|c||}
\hline \hline
Sec&$N_R$&$\psi$&$x^{12}$&$x^{34}$&$x^{56}$&$\overline{\psi}^{~1,...,5}$&$\overline{\eta}^{~1}$&$\overline{\eta}^{~2}$&$\overline{\eta}^{~3}$&$\overline{\phi}^{~1,...,8}$\\
\hline \hline
$\vec{v}$&3&1&0&0&0&$\frac{2}{3}$,...,$\frac{2}{3}$&$\frac{2}{3}$&$\frac{2}{3}$&$\frac{2}{3}$&0,...,0,$\frac{2}{3}$, $\frac{2}{3}$, $\frac{2}{3}$\\
\hline \hline
\end{tabular}
\vspace{5 mm}\\
\begin{tabular}{||c|c|c|c|c||}
\hline \hline
Sec&O&$y^{~1,2}w^{~5,6}||\overline{y}^{~1,2}\overline{w}^{~5,6}$&$y^{~3,...,6}||\overline{y}^{~3,...,6}$&$w^{~1,...,4}||\overline{w}^{~1,...,4}$\\
\hline \hline
$\vec{v}$&3&0,0,1,1$||$0,0,$\frac{2}{3}$, $\frac{2}{3}$&0,0,0,0$||$0,0,$\frac{2}{3}$,$\frac{2}{3}$&0,0,1,1$||$$\frac{2}{3}$,$\frac{2}{3}$,$\frac{2}{3}$,$\frac{2}{3}$\\
\hline \hline
\end{tabular}
\vspace{3 mm}\\
$k_{\vec{v},j}$ = (0, 1, 0, 0, 0)
\label{tab: 3G_LRSym_NAHE_O3L1_Example_BVs}
\end{center}
\end{table}
\begin{table}
\caption{The particle content of the three-generation Left-Right Symmetric Model. This model also has 7 $U(1)$'s and $N=1$ ST SUSY.}
\begin{center}
\begin{tabular}{||c|c|c|c|c|c|c||}
\hline \hline
\textbf{QTY}&$SU(2)_L$&$SU(2)_R$&$SU(3)_C$&$SU(3)$&$SU(5)$&$SO(10)$\\
\hline \hline
1&$2_{ }$&$2_{ }$&$\overline{3}$&$1_{ }$&$1_{ }$&$1_{ }$\\
\hline
2&$2_{ }$&$1_{ }$&$3_{ }$&$1_{ }$&$1_{ }$&$1_{ }$\\
\hline
2&$2_{ }$&$1_{ }$&$1_{ }$&$3_{ }$&$1_{ }$&$1_{ }$\\
\hline
3&$2_{ }$&$1_{ }$&$1_{ }$&$1_{ }$&$1_{ }$&$1_{ }$\\
\hline
3&$2_{ }$&$1_{ }$&$1_{ }$&$1_{ }$&$5_{ }$&$1_{ }$\\
\hline
1&$2_{ }$&$1_{ }$&$1_{ }$&$\overline{3}$&$1_{ }$&$1_{ }$\\
\hline
1&$2_{ }$&$1_{ }$&$\overline{3}$&$1_{ }$&$1_{ }$&$1_{ }$\\
\hline
2&$1_{ }$&$2_{ }$&$1_{ }$&$3_{ }$&$1_{ }$&$1_{ }$\\
\hline
2&$1_{ }$&$2_{ }$&$1_{ }$&$1_{ }$&$\overline{5}$&$1_{ }$\\
\hline
3&$1_{ }$&$2_{ }$&$1_{ }$&$1_{ }$&$1_{ }$&$1_{ }$\\
\hline
1&$1_{ }$&$2_{ }$&$1_{ }$&$1_{ }$&$5_{ }$&$1_{ }$\\
\hline
1&$1_{ }$&$2_{ }$&$1_{ }$&$\overline{3}$&$1_{ }$&$1_{ }$\\
\hline
3&$1_{ }$&$2_{ }$&$\overline{3}$&$1_{ }$&$1_{ }$&$1_{ }$\\
\hline
2&$1_{ }$&$1_{ }$&$3_{ }$&$3_{ }$&$1_{ }$&$1_{ }$\\
\hline
4&$1_{ }$&$1_{ }$&$3_{ }$&$1_{ }$&$1_{ }$&$1_{ }$\\
\hline
2&$1_{ }$&$1_{ }$&$3_{ }$&$\overline{3}$&$1_{ }$&$1_{ }$\\
\hline
3&$1_{ }$&$1_{ }$&$1_{ }$&$3_{ }$&$1_{ }$&$1_{ }$\\
\hline
2&$1_{ }$&$1_{ }$&$1_{ }$&$1_{ }$&$\overline{10}$&$1_{ }$\\
\hline
1&$1_{ }$&$1_{ }$&$1_{ }$&$1_{ }$&$\overline{5}$&$1_{ }$\\
\hline
1&$1_{ }$&$1_{ }$&$1_{ }$&$1_{ }$&$1_{ }$&$\overline{16}$\\
\hline
3&$1_{ }$&$1_{ }$&$1_{ }$&$1_{ }$&$1_{ }$&$10_{ }$\\
\hline
9&$1_{ }$&$1_{ }$&$1_{ }$&$1_{ }$&$1_{ }$&$1_{ }$\\
\hline
3&$1_{ }$&$1_{ }$&$1_{ }$&$1_{ }$&$1_{ }$&$16_{ }$\\
\hline
5&$1_{ }$&$1_{ }$&$1_{ }$&$1_{ }$&$5_{ }$&$1_{ }$\\
\hline
1&$1_{ }$&$1_{ }$&$1_{ }$&$\overline{3}$&$\overline{10}$&$1_{ }$\\
\hline
1&$1_{ }$&$1_{ }$&$\overline{3}$&$3_{ }$&$1_{ }$&$1_{ }$\\
\hline
1&$1_{ }$&$1_{ }$&$\overline{3}$&$1_{ }$&$\overline{5}$&$1_{ }$\\
\hline \hline
\end{tabular}
\label{tab: 3G_LRSym_NAHE_O3L1_Example_Particles}
\end{center}
\end{table}
\begin{table}
\caption{The observable matter content of the three-generation Left-Right Symmetric Model.}
\begin{center}
\begin{tabular}{||c|c|c|c|c|c|c||}
\hline \hline
\textbf{QTY}&$SU(2)_L$&$SU(2)_R$&$SU(3)_C$&$SU(3)$&$SU(5)$&$SO(10)$\\
\hline \hline
1&$2_{ }$&$2_{ }$&$\overline{3}$&$1_{ }$&$1_{ }$&$1_{ }$\\
\hline
2&$2_{ }$&$1_{ }$&$3_{ }$&$1_{ }$&$1_{ }$&$1_{ }$\\
\hline
2&$2_{ }$&$1_{ }$&$1_{ }$&$3_{ }$&$1_{ }$&$1_{ }$\\
\hline
3&$2_{ }$&$1_{ }$&$1_{ }$&$1_{ }$&$1_{ }$&$1_{ }$\\
\hline
3&$2_{ }$&$1_{ }$&$1_{ }$&$1_{ }$&$5_{ }$&$1_{ }$\\
\hline
1&$2_{ }$&$1_{ }$&$1_{ }$&$\overline{3}$&$1_{ }$&$1_{ }$\\
\hline
1&$2_{ }$&$1_{ }$&$\overline{3}$&$1_{ }$&$1_{ }$&$1_{ }$\\
\hline
2&$1_{ }$&$2_{ }$&$1_{ }$&$3_{ }$&$1_{ }$&$1_{ }$\\
\hline
2&$1_{ }$&$2_{ }$&$1_{ }$&$1_{ }$&$\overline{5}$&$1_{ }$\\
\hline
3&$1_{ }$&$2_{ }$&$1_{ }$&$1_{ }$&$1_{ }$&$1_{ }$\\
\hline
1&$1_{ }$&$2_{ }$&$1_{ }$&$1_{ }$&$5_{ }$&$1_{ }$\\
\hline
1&$1_{ }$&$2_{ }$&$1_{ }$&$\overline{3}$&$1_{ }$&$1_{ }$\\
\hline
3&$1_{ }$&$2_{ }$&$\overline{3}$&$1_{ }$&$1_{ }$&$1_{ }$\\
\hline
2&$1_{ }$&$1_{ }$&$3_{ }$&$3_{ }$&$1_{ }$&$1_{ }$\\
\hline
4&$1_{ }$&$1_{ }$&$3_{ }$&$1_{ }$&$1_{ }$&$1_{ }$\\
\hline
2&$1_{ }$&$1_{ }$&$3_{ }$&$\overline{3}$&$1_{ }$&$1_{ }$\\
\hline
1&$1_{ }$&$1_{ }$&$\overline{3}$&$3_{ }$&$1_{ }$&$1_{ }$\\
\hline
1&$1_{ }$&$1_{ }$&$\overline{3}$&$1_{ }$&$\overline{5}$&$1_{ }$\\
\hline \hline
\end{tabular}
\label{tab: 3G_LRSym_NAHE_O3L1_Example_Observable_Matter}
\end{center}
\end{table}
This model has three net generations of quarks, but no net generations of anti-quarks.
Additionally there are thirty left- and right-handed lepton doublets.
Other exotics include a quark triplet with left- and right-handed isospin, eight quark and ten anti-quark triplets without isospin.
Thus, this model is not a favorable candidate for a quasi-realistic three-generation model.
It does serve as a proof of concept that three generation models can be built with single-layer extensions to the NAHE set, however.
\section{Conclusions}\label{sec: NAHE_Conclusions}
The statistics presented in this paper make it clear that the NAHE set does serve its intended purpose as a basis for quasi-realistic WCFFHS models at a statistical level.
Three generation models were constructed from order-3 extensions to the NAHE set.
A summary of the GUT group analysis is presented in Table \ref{tab: NAHE_GUT_Generations}.
\begin{table}
\begin{center}
\caption{A summary of the GUT group study with regard to the number of chiral fermion generations in the NAHE set investigation.}
\begin{tabular}{||c|c|c||}
\hline \hline
GUT&Net Chiral Generations?&Three Generations?\\
\hline \hline
O2L1 $SO(10)$&Yes&No\\
\hline
O2L1 Pati-Salam&No&No\\
\hline
O3L1 $E_6$&Yes&No\\
\hline
O3L1 $SO(10)$&Yes&No\\
\hline
O3L1 $SU(5)\otimes U(1)$&Yes&Yes\\
\hline
O3L1 Pati-Salam&Yes&No\\
\hline
O3L1 L-R Symmetric&Yes&Yes\\
\hline
O3L1 MSSM&Yes&Yes\\
\hline \hline
\end{tabular}
\label{tab: NAHE_GUT_Generations}
\end{center}
\end{table}
Two three-generation models were discussed - a flipped-$SU(5)$ model and a Left-Right Symmetric model.
While they did have the requisite number of chiral matter generations, there were several unfavorable properties in both models that prevented them from being considered quasi-realistic.
They are a proof that three-generation models with geometric interpretations can be built with order-3 basis vector extensions.

The distributions of ST SUSYs across the GUT group subsets remained largely the same, save the $E_6$ models, which displayed a greater statistical tendency for ST SUSY enhancements.
Further examination revealed that all of the exceptional groups, as well as $SU(9)$ and $SU(12)$, display more enhanced ST SUSY models for the NAHE + O3L1 data set. 
It was also shown that the presence of the $\vec{S}$ did not significantly impact the gauge content.
However, the matter content of the models without $\vec{S}$ is affected, as the $\vec{S}$ sector produces states other than SUSY partners.
\section{Acknowledgements}
This work was supported by funding from Baylor University.

\end{document}